\documentclass[12pt]{article}
\usepackage[height=8.5in,width=6.4in]{geometry}

\usepackage{xparse}
\usepackage{amssymb, amsmath}
\usepackage{xcolor}
\usepackage{tikz}
\usepackage{cite}
\usepackage[linktocpage]{hyperref}
\usepackage{mathtools}
\usepackage[vcentermath]{youngtab}

\DeclareFontShape{OT1}{cmr}{mx}{n}%
    {<->cmr10}{}
\renewcommand{\hat}{\widehat}
\renewcommand{\tilde}{\widetilde}

\numberwithin{equation}{section}

\begin{document}
\setcounter{tocdepth}{2}

\begin{titlepage}
\begin{flushright} 
YITP-20-164
\end{flushright}

\vskip 1cm

\begin{center}

{\LARGE\fontseries{mx}\selectfont
Argyres-Douglas Theories, S-duality and \\
 AGT Correspondence
\par
}

\vskip 1.3cm

Takuya Kimura,$^{\diamondsuit, 1}$ Takahiro Nishinaka,$^{\clubsuit, 1}$ Yuji Sugawara,$^{\heartsuit, 1}$ and Takahiro Uetoko$^{\spadesuit, 2}$

\vskip .7cm

{\it
$^1$ Department of Physical Sciences, College of Science and Engineering\\
 Ritsumeikan University, Shiga 525-8577, Japan
}

\vskip.4cm

{\it
$^2$ Center for Gravitational Physics, Yukawa Institute for Theoretical Physics\\
 Kyoto University, Kyoto 606-8502, Japan
}

\end{center}

\vskip.5cm

\begin{abstract}
We propose a Nekrasov-type formula for the instanton partition functions of four-dimensional $\mathcal{N}=2\;\,U(2)$
  gauge theories coupled to
 $(A_1,D_{2n})$ Argyres-Douglas theories. This is carried out by extending the
 generalized AGT correspondence to the case of $U(2)$ gauge group, which
 requires us to define irregular states of
 the direct sum of Virasoro and Heisenberg algebras. Using our formula,
 one can evaluate the contribution of the $(A_1,D_{2n})$ theory at
 each fixed point on the $U(2)$ instanton moduli space.
As an application, we evaluate the instanton partition function of the
 $(A_3,A_3)$ theory to find it in a peculiar relation to that of $SU(2)$
 gauge theory with four fundamental flavors. From this relation, we read
 off how the S-duality group acts on the UV gauge coupling of the
 $(A_3,A_3)$ theory. 

\end{abstract}

\renewcommand{\thefootnote}{\fnsymbol{footnote}}
\footnotetext[0]{
\!\!\!\!\!\!\!\!\!\!\!$^{\diamondsuit}$rp0047ir@ed.ritsumei.ac.jp,
\\
$^{\clubsuit}$nishinak@fc.ritsumei.ac.jp,
\\
$^{\heartsuit}$ysugawa@se.ritsumei.ac.jp,
\\
$^{\spadesuit}$takahiro.uetoko@yukawa.kyoto-u.ac.jp}
\renewcommand{\thefootnote}{\arabic{footnote}}

\end{titlepage}

\tableofcontents

\section{Introduction}
\label{sec:intro}

The Nekrasov partition function of four-dimensional $\mathcal{N}=2$ gauge
theories with Lagrangian description can often be exactly evaluated by
supersymmetric localization \cite{Nekrasov:2002qd,Nekrasov:2003rj}.
The result is 
%generally
written as a sum over fixed points of a torus action on the instanton moduli
space. 
% When the gauge group is $U(N)$, this sum can be regarded as a sum
% over $N$-tuples of Young diagrams, since the fixed points in this case are in
% one-to-one correspondence with such $N$-tuples.
 For instance, for $U(2)$ gauge theory with $N_f$ fundamental
hypermultiplets, the partition function is evaluated as
\begin{align}
 \mathcal{Z}_{U(2)}^{N_f} =
 \mathcal{Z}_\text{pert}\sum_{Y_1,Y_2}\Lambda^{b_0(|Y_1|+|Y_2|)}\mathcal{Z}^\text{vec}_{Y_1,Y_2}(a)
 \prod_{i=1}^{N_f} \mathcal{Z}^\text{fund}_{Y_1,Y_2}(a,m_i)~,
\label{eq:Nek1}
\end{align}
where $b_0 \equiv 4-N_f$, $\Lambda$ is a dynamical scale, $a$ is the
vacuum expectation value (VEV) of a scalar, $Y_k$ are Young diagrams,
and $|Y_k|$ is the number of boxes in $Y_k$. The sum over $(Y_1,Y_2)$ can be
regarded as a sum over fixed points on the moduli space of $U(2)$
instantons, and
$\mathcal{Z}^\text{vec}_{Y_1,Y_2}$ and
$\mathcal{Z}^\text{fund}_{Y_1,Y_2}$ are respectively the contributions
of the gauge and matter sectors at these fixed points. The prefactor,
$\mathcal{Z}_\text{pert}$, is the perturbative contribution that makes the power series in $\Lambda$ start with $1$. 
%The above exact expression for the partition function
%has revealed various non-perturbative phenomena of the theory.

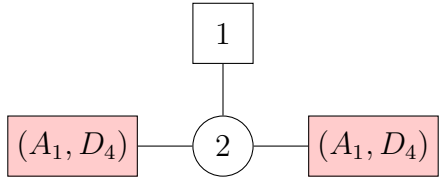
\begin{figure}
 \begin{center}
\begin{tikzpicture}[gauge/.style={circle,draw=black,inner sep=0pt,minimum size=8mm},flavor/.style={rectangle,draw=black,inner sep=0pt,minimum size=8mm},AD/.style={rectangle,draw=black,fill=red!20,inner sep=0pt,minimum size=8mm},auto]

\node[AD] (1) at (-2,0) {\;$(A_1,D_{4})$\;};
\node[gauge] (2) at (0,0) [shape=circle] {\;$2$\;} edge (1);
\node[AD] (3) at (2,0)  {\;$(A_1,D_{4})$\;} edge (2);
\node[flavor] (4) at (0,1.5) {\;1\;} edge (2);

  \end{tikzpicture}
\caption{The quiver diagram of the $(A_3,A_3)$ AD theory. The left and right
  boxes stand for two copies of $(A_1,D_4)$ theory while the top box
  stands for a fundamental hypermultiplet of $SU(2)$. The middle circle
  stands for $SU(2)$ gauge group coupled to these ``matters.''}
\label{fig:quiver1}
 \end{center}
\end{figure}

Despite the above success for Lagrangian theories, there is a rich class of four-dimensional $\mathcal{N}=2$
gauge theories whose partition functions are still to be evaluated. These
theories involve strongly-coupled CFTs in their matter sector, and therefore
their partition functions cannot be directly evaluated by supersymmetric
localization.
Among other theories, conformal gauge theories coupled to
Argyres-Douglas (AD) theories are of particular
importance in this class, since they provide a new class of
$\mathcal{N}=2$ S-dualities \cite{Buican:2014hfa, DelZotto:2015rca, Cecotti:2015hca, Xie:2016uqq, Xie:2017vaf,
Buican:2017fiq, Xie:2017aqx, Buican:2018ddk}. We call such conformal gauge
theories ``conformally gauged AD theories.'' 
One of the simplest examples of such theories is the $(A_3,A_3)$ theory
described by the quiver
diagram in Fig.~\ref{fig:quiver1}, where the $(A_1,D_4)$ theory is a
particular AD theory.\footnote{This AD theory is also called $H_2$
theory, $D_2(SU(3))$ theory, and $(A_2,A_2)$ theory in the
literature. The first series of papers on AD theories are
\cite{Argyres:1995jj, Argyres:1995xn, Eguchi:1996vu}.}
The beta function of the gauge coupling vanishes here since the
contributions of the gauge and matter sectors are exactly canceled.
%We call these conformal gauge theories coupled to AD theories ``conformally
%gauged AD theories.'' 

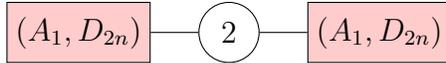
\begin{figure}
 \begin{center}
\begin{tikzpicture}[gauge/.style={circle,draw=black,inner sep=0pt,minimum size=8mm},flavor/.style={rectangle,draw=black,inner sep=0pt,minimum size=8mm},AD/.style={rectangle,draw=black,fill=red!20,inner sep=0pt,minimum size=8mm},auto]

\node[AD] (1) at (-2,0) {\;$(A_1,D_{2n})$\;};
\node[gauge] (2) at (0,0) [shape=circle] {\;$2$\;} edge (1);
\node[AD] (3) at (2,0)  {\;$(A_1,D_{2n})$\;} edge (2);

  \end{tikzpicture}
\caption{A typical example of gauge theories studied in the generalized
  AGT correspondence, where $n$ is a positive integer. The circle stands for $SU(2)$ gauge group, and each
box stands for an $(A_1,D_{2n})$ theory. The $SU(2)$ vector multiplet
is  diagonally gauging an $SU(2)$ sub-group of the $(A_1,D_{2n})$
  theories.}
\label{fig:quiver2}
 \end{center}
\end{figure}

While the partition functions of conformally gauged AD theories have
not been evaluated, there exists a series of {\it non-conformally}
gauged AD theories whose
partition functions were evaluated via a generalization
\cite{Bonelli:2011aa, Gaiotto:2012sf} of the AGT
correspondence \cite{Alday:2009aq, Gaiotto:2009ma} (See
\cite{Kanno:2012xt, Nishinaka:2012kn, Rim:2012tf, Kanno:2013vi, Matsuo:2014rba, 
 Choi:2015idw,  Rim:2016msx,
Itoyama:2018wbh, Itoyama:2018gnh,Nishinaka:2019nuy} for
recent developments on this generalization). In particular, for the theory described by the quiver in Fig.~\ref{fig:quiver2}, the partition function was
evaluated as the inner product of so-called ``irregular states'' of
Virasoro algebra. The application of the generalized AGT
correspondence is possible here 
%shown in Fig.~\ref{fig:quiver2} 
since these theories 
can be engineered by compactifying the 6d $(2,0)$ $A_1$
theory on a Riemann surface.

The purpose of this paper is to propose a way to compute the partition
function of the conformally gauged AD theory 
%shown 
in
Fig.~\ref{fig:quiver1}, using the
generalized AGT correspondence for the non-conformally gauged AD
theory
%shown 
in Fig.~\ref{fig:quiver2}. Since the former has no known construction
from the 
6d (2,0) $A_1$ theory, one cannot directly apply the AGT correspondence
to it.\footnote{The theory shown in Fig.~\ref{fig:quiver1} can be
constructed by compactifying the 6d (2,0) $A_2$ or $A_3$ theory on a Riemann
surface \cite{Xie:2012hs, Beem:2020pry}. This suggests a possibility of studying its partition function via the
higher-rank generalization of the AGT correspondence
\cite{Wyllard:2009hg}. Going in this direction would, however, be
non-trivial and involved since the 2d
side is now a higher-rank Toda theory. In this paper, we take a
different route.} Instead, our strategy is to apply the generalized AGT correspondence to
the latter (with a
small but crucial modification discussed below) and decompose the resulting
partition function as a sum over Young diagrams, as in
\eqref{eq:Nek1}.
% for Lagrangian theories. 
While the theory has a strongly-coupled matter sector, such a
decomposition is expected, since the gauge sector of the theory is still
described by Lagrangian and its path integral is expected to
lead to a sum
over fixed points on the instanton moduli space.
Once such a decomposition is obtained, 
one can read off the contribution of the $(A_1,D_{2n})$ theory at each
%torus 
fixed point, say $\mathcal{Z}^{(A_1,D_{2n})}_{Y_1,Y_2}$.
% on the instanton moduli space. 
%Given an expression for
%$\mathcal{Z}^{(A_1,D_{2n})}_{Y_1,Y_2}$, 
It is then
straightforward to compute the 
%instanton 
partition function of the
conformally gauged AD theory shown in Fig.~\ref{fig:quiver1}.

One important point in the above discussion is that the decomposition
of the form \eqref{eq:Nek1} is for $U(2)$ gauge group while the
generalized AGT correspondence is for $SU(2)$ 
%(non-conformal) 
gauge
group. Indeed, for fixed points on the instanton moduli space to be
labeled by $(Y_1,Y_2)$, the gauge group must be $U(2)$ instead of $SU(2)$.
Therefore, to decompose the resulting partition function as a sum
over $(Y_1,Y_2)$, 
%to read off the contribution
%$\mathcal{Z}_{Y_1,Y_2}^{(A_1,D_4)}$ with the above strategy,
 we need a $U(2)$ version of the generalized AGT
correspondence.
% where the gauge group is
%replaced with $U(2)$ in Fig.~\ref{fig:quiver2}. 
For the original AGT correspondence, such a $U(2)$
version is realized by considering the direct sum of Virasoro and
Heisenberg algebras ($Vir\oplus H$) on the two-dimensional side \cite{Alba:2010qc}. In this
paper, we extend it to the generalized AGT
correspondence, by considering irregular states of $Vir\oplus H$.

Given the $U(2)$ version of the generalized AGT correspondence, one can
easily 
%write down the partition function of $U(2)$ gauge theory
%coupled to two copies of $(A_1,D_{2n})$ as the inner
%product of irregular states of $Vir\times H$. Using a nice basis of
%highest weight modules of
%$Vir\times H$ \cite{Alba:***}, one can 
decompose the partition function of $U(2)$ gauge theory coupled to two
$(A_1,D_{2n})$ as
% \begin{align}
%  \mathcal{Z}_{U(2)}^{2\times (A_1,D_{2n})} =
%  \sum_{Y_1,Y_2}\Lambda^{\beta(|Y_1|+|Y_2|)}\mathcal{Z}^\text{vec}_{Y_1,Y_2}(a)
%  \prod_{i=1}^2\mathcal{Z}_{Y_1,Y_2}^{(A_1,D_{2n})}(a,m_i,\vec{d}_i,\vec{u}_i)~,
% \label{eq:Nek2}
% \end{align}
\begin{align}
 \mathcal{Z}_{U(2)}^{2\times (A_1,D_{2n})} =
 \mathcal{Z}_\text{pert}\sum_{Y_1,Y_2}\Lambda^{b_0(|Y_1|+|Y_2|)}\mathcal{Z}^\text{vec}_{Y_1,Y_2}(a)
 \mathcal{Z}_{Y_1,Y_2}^{(A_1,D_{2n})}(a,m,\pmb{d},\pmb{u})\tilde{\mathcal{Z}}_{Y_1,Y_2}^{(A_1,D_{2n})}(a,\tilde{m},\tilde{\pmb{d}},\tilde{\pmb{u}})~,
\label{eq:Nek2}
\end{align}
where $m,\pmb{d}=(d_1,\cdots,d_{n-1})$ and $\pmb{u}=(u_1,\cdots,u_{n-1})$ are respectively the mass,
relevant couplings and
VEVs of Coulomb branch operators of the $(A_1,D_{2n})$ theory, and
$b_0\equiv 2/n$ is the coefficient of the one-loop $\beta$ function.
We interpret the sum over $(Y_1,Y_2)$ as a sum over fixed
points on the moduli space of $U(2)$ instantons, and identify
 $\mathcal{Z}^{(A_1,D_{2n})}_{Y_1,Y_2}$ and
 $\tilde{\mathcal{Z}}^{(A_1, D_{2n})}_{Y_1,Y_2}$ as contributions from
the $(A_1,D_{2n})$ theories at each fixed point. The difference between  $\mathcal{Z}^{(A_1,D_{2n})}_{Y_1,Y_2}$ and
 $\tilde{\mathcal{Z}}^{(A_1, D_{2n})}_{Y_1,Y_2}$ is interpreted as coming
 from how
 the $U(1)$ part of the gauge group is coupled to $(A_1,D_{2n})$.

With the above identification of  $\mathcal{Z}^{(A_1,D_{2n})}_{Y_1,Y_2}$,
we evaluate the partition function of the $(A_3,A_3)$ theory as
follows. We start with the quiver description shown in
Fig.~\ref{fig:quiver1}. When the gauge group in the quiver is $U(2)$, the partition function is
evaluated as
\begin{align}
 \mathcal{Z}_{U(2)} = \mathcal{Z}_\text{pert}
 \sum_{Y_1,Y_2}q^{|Y_1|+|Y_2|}\mathcal{Z}^\text{vec}_{Y_1,Y_2}(a)\mathcal{Z}^\text{fund}_{Y_1,Y_2}(a,M)
\prod_{i=1}^2\mathcal{Z}_{Y_1,Y_2}^{(A_1,D_4)}(a,m_i,d_i,u_i)~,
\label{eq:Nek3}
\end{align}
where $q$ is the exponential of the marginal gauge coupling, and $M$ is
the mass of the hypermultiplet. The prefactor $\mathcal{Z}_\text{pert}$
is again the perturbative contribution that makes the power series in
$q$ start with $1$. Since
 $\mathcal{Z}^{(A_1,D_{2n})}_{Y_1,Y_2}$ is already read off from the decomposition
\eqref{eq:Nek2}, one can explicitly compute the above partition
function. 
When the gauge group is $SU(2)$, the partition function differs from
\eqref{eq:Nek3} by the contribution of the $U(1)$-part of the gauge group. Indeed, according to a general
discussion in \cite{Alday:2009aq}, the partition function for $SU(2)$
gauge group is expected to be given by
\begin{align}
 \mathcal{Z}_{SU(2)} =
 \frac{\mathcal{Z}_{U(2)}}{\mathcal{Z}_{U(1)}}~,
\label{eq:U1}
\end{align}
where $\mathcal{Z}_{U(1)}$ is  the partition
function of the $U(1)$ part, and called ``$U(1)$-factor.''
%Since $a$ is the VEV of a scalar neutral under $U(1)$, the $U(1)$-factor
%is expected to be independent of $a$. This implies that, when all the massive deformations but
%$a$ are turned off, $\mathcal{Z}_{U(1)} = 1$ and therefore
%$\mathcal{Z}_{SU(2)} = \mathcal{Z}_{U(2)}$. 
We use \eqref{eq:Nek3} and
\eqref{eq:U1} to show in particular that the S-duality of the $(A_3,A_3)$ theory is in
a peculiar relation to that of $SU(2)$ gauge theory with $4$ fundamental
flavors. It is an interesting open problem to see how this peculiar
relation is connected to a similar relation between the Schur index of
the same pair of theories discussed in \cite{Buican:2019kba}.

The organization of the rest of this paper is the following. In
Sec.~\ref{sec:AGT}, we briefly review the AGT correspondence and its
generalization to AD theories. In Sec.~\ref{sec:V+H}, we consider a
$U(2)$-version of the generalized AGT correspondence, in terms of
irregular states of $Vir\oplus H$. In Sec.~\ref{sec:Nek-AD}, we derive
a formula for $\mathcal{Z}_{Y_1,Y_2}^{(A_1,D_{2n})}$ corresponding to
the gauged $(A_1,D_{2n})$ theory, using the $U(2)$-version of the
generalized AGT correspondence discussed in the previous section. In
Sec.~\ref{sec:A3A3}, we evaluate the partition function of $(A_3,A_3)$
theory using our formula for
$\mathcal{Z}_{Y_1,Y_2}^{(A_1,D_{4})}$. We particularly discuss the
S-duality of the theory in connection to the S-duality of $SU(2)$ gauge
theory with four flavors. In Sec.~\ref{sec:Conclusions}, we conclude and
discuss future directions. There are several
appendices. Sec.~\ref{app:Nek} includes Nekrasov's formulae for
Lagrangian sectors. Sec.~\ref{app:basis} contains the first few examples
of states in a special basis of the highest weight module of $Vir\oplus H$ that we
will use in Sec.~\ref{sec:Nek-AD}. In Sec.~\ref{app:F2}, we explain how
the prepotential of the $(A_3,A_3)$ theory is constrained by the
invariance under \eqref{eq:T-massive} when all the
massive deformations are turned on.

\section{Generalized AGT correspondence}
\label{sec:AGT}

 In this section, we briefly review the AGT correspondence
 \cite{Alday:2009aq, Gaiotto:2009ma} and its
 generalization \cite{Bonelli:2011aa,
 Gaiotto:2012sf}.
Suppose that $\mathcal{T}_{\mathcal{C}}$ is a four-dimensional $\mathcal{N}=2$
superconformal field theory obtained by compactifying the 6d $(2,0)\; A_1$ theory on a punctured Riemann surface $\mathcal{C}$. It is known
that punctures on $\mathcal{C}$ can be ``regular'' or ``irregular''
depending on whether a six-dimensional BPS scalar operator has a
simple pole or higher order pole at it. When the BPS operator
has a pole of order $(n+1)$ at an irregular puncture, we say the
puncture is of rank $n$. While this rank can be an integer or half-integer
in general, we only consider integer ranks in this paper.

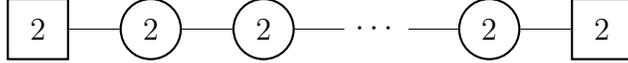
\begin{figure}
 \begin{center}
\begin{tikzpicture}[gauge/.style={circle,draw=black,thick,inner sep=0pt,minimum size=8mm},flavor/.style={rectangle,draw=black,thick,inner sep=0pt,minimum size=8mm},auto]

\node[flavor] (1) at (-8,0) {\;$2$\;};
\node[gauge] (2) at (-6.5,0) [shape=circle] {\;$2$\;} edge (1);
\node[gauge] (3) at (-5,0) [shape=circle] {\;$2$\;} edge (2);
\node (4) at (-3.5,0) {$\cdots$} edge (3);
\node[gauge] (5) at (-2,0) {\;$2$\;} edge (4);
\node[flavor] (6) at (-.5,0) {\;$2$\;} edge (5);

  \end{tikzpicture}
\caption{The quiver diagram of the class $\mathcal{S}$ theory
  $\mathcal{T}_{\mathcal{C}}$ for $\mathcal{C}$ being sphere with $(n+2)$
  regular punctures. There are $(n-1)$ circles, each of which stands for an $SU(2)$ gauge
  group.}
\label{fig:quiver0}
 \end{center}
\end{figure}

Let us first focus on the case in which $\mathcal{C}$ has no irregular
puncture. In this case, the AGT
correspondence \cite{Alday:2009aq} implies that the Nekrasov partition function of
$\mathcal{T}_{\mathcal{C}}$ is identical to the conformal block of
Liouville theory on $\mathcal{C}$, where the Liouville charge $Q$ is
related to the $\Omega$-background parameters by
$Q=(\epsilon_1+\epsilon_2)/\sqrt{\epsilon_1\epsilon_2}$. As an example,
let us consider the case of $\mathcal{C}$ being
a sphere with $(n+2)$ regular punctures. The corresponding  $\mathcal{T}_{\mathcal{C}}$ is
a linear quiver $SU(2)$ gauge theory as shown in
Fig.~\ref{fig:quiver0}. The AGT correspondence implies
\begin{align}
 \mathcal{Z}_{SU(2)}(\vec{a};m_1,\cdots,m_{n+2}) =
 \mathcal{F}_{\alpha_{0}}{}^{\alpha_{1}}{}_{\beta_1}{}^{\alpha_2}{}_{\beta_2}{}^{\alpha_3}\cdots
 {}_{\beta_{n-1}}{}^{\alpha_{n}}{}_{\alpha_{n+1}}~,
\label{eq:linear-quiver}
\end{align}
where the LHS is the Nekrasov partition function of $\mathcal{T}_{C}$
with $\vec{a} \equiv(a_1,\cdots,a_{n-1})$ being the VEVs of Coulomb branch operators, and $m_i$ being
the masses of fundamental and bi-fundamental hypermultiplets. The
subscript, $SU(2)$, is for emphasizing that
the gauge group of $\mathcal{T}_{C}$ is $SU(2)^{n-1}$. The
RHS of \eqref{eq:linear-quiver} is the $(n+2)$-point conformal block of Liouville
theory with $\beta_i$ being intermediate momenta, and $\alpha_{i}$
being external momenta. The 4d and 2d parameters are
related by
\begin{align}
 \frac{a_i}{\sqrt{\epsilon_1 \epsilon_2}} &= \beta_i +
 \frac{Q}{2}~,\qquad \frac{m_i}{\sqrt{\epsilon_1\epsilon_2}} = \alpha_i + \frac{Q}{2}~.
\end{align}
% where we defined $\hbar\equiv \sqrt{-\epsilon_1\epsilon_2}$, $\alpha_0 \equiv (\tilde{\alpha}_0 -
% \tilde{\alpha}_1)/2,\,\alpha_1 \equiv (\tilde{\alpha}_0 +
% \tilde{\alpha}_1)/2,\,\alpha_n \equiv (\tilde{\alpha}_n +
% \tilde{\alpha}_{n+1})/2$ and $\alpha_{n+1} \equiv
% (\tilde{\alpha}_n-\tilde{\alpha}_{n+1})/2$.
 In the rest of this paper,
we rescale dimensionful parameters so that
$\epsilon_1\epsilon_2=1$, as in  \cite{Alday:2009aq}, and write things
in terms of the 2d parameters. When we
need to recover the full $\epsilon_i$-dependence, we rescale the 2d
parameters as
\begin{align}
 \alpha_i\to \frac{\alpha_i}{\sqrt{\epsilon_1\epsilon_2}}~,\qquad \beta_i\to \frac{\beta_i}{\sqrt{\epsilon_1\epsilon_2}}~,\qquad Q \to
 \frac{Q}{\sqrt{\epsilon_1\epsilon_2}}~.
\label{eq:recover}
\end{align}

The AGT correspondence has been generalized to the case in which
$\mathcal{C}$ has irregular punctures \cite{Gaiotto:2009ma,Bonelli:2011aa,
Gaiotto:2012sf}.
The most important difference from the original AGT is that
$\mathcal{T}_{\mathcal{C}}$ generically involves an AD theory \cite{Bonelli:2011aa,
Gaiotto:2012sf}. One simple example is the case in which
$\mathcal{C}$ is a sphere with one irregular puncture of rank $n$ and
one regular puncture. In this case, $\mathcal{T}_{\mathcal{C}}$ is an AD theory called
the $(A_1,D_{2n})$ theory, and its partition function is identified as the two-point function of 2d operators
corresponding to the punctures on $\mathcal{C}$. By the state-operator map, this two-point function
 is mapped to the inner product of two states
\begin{align}
 \mathcal{Z}_{(A_1,D_{2n})} = \langle a|I^{(n)}\rangle~,
\label{eq:ZAD1}
\end{align}
where $|a\rangle$ is the Virasoro primary state corresponding to the
regular puncture, while $|I^{(n)}\rangle$ is a state corresponding to
the rank-$n$ irregular puncture.

A more interesting situation arises when
 $\mathcal{C}$ has two
 irregular punctures of rank $n$. In this case, $\mathcal{T}_{\mathcal{C}}$ is no longer
 conformal, but is an $SU(2)$ gauge theory coupled to two
 copies of $(A_1,D_{2n})$ theories. 
% This generalization has two important differences compared to the original AGT correspondence.
%First, in the presence of an irregular puncture, the four-dimensional
%theory $\mathcal{T}_{\mathcal{C}}$ is not necessarily
%conformal. The second difference is that, even though the gauge sector
%of $\mathcal{T}_{\mathcal{C}}$ is still composed of $SU(2)$ vector
%multiplets, the matter sector of $\mathcal{T}_{\mathcal{C}}$ generically
%contains AD
%theories \cite{Bonelli:2011aa,
%Gaiotto:2012sf}. For example, when $\mathcal{C}$ has two irregular
%singularities of rank $n$ without regular ones, $\mathcal{T}_{\mathcal{C}}$ is
%$SU(2)$ gauge theory coupled to two copies of $(A_1,D_{2n})$
%theories. 
The quiver diagram of this gauge theory is shown in
Fig.~\ref{fig:quiver2}.
Here, each $(A_1,D_{2n})$ in the matter sector is
 associated with an irregular puncture on $\mathcal{C}$. The
 $\beta$-function coefficient of this $SU(2)$ gauge coupling is
 evaluated as $\beta_0 = 2/n$, and therefore $\mathcal{T}_{\mathcal{C}}$
 is never conformal.
The generalized AGT correspondence then implies that the partition
function of this theory is given by
% $\mathcal{T}_{\mathcal{C}}$ is identical to the two point
% function of some operators in the Liouville CFT that correspond to two
% irregular singularities on $\mathcal{C}$. Using the conformal invariance
% in two dimensions, this two-point function is mapped to the inner
% product of some states, i.e.,
\begin{align}
\mathcal{Z}_{SU(2)}^{2\times (A_1,D_{2n})} &= \langle I^{(n)}|I^{(n)}\rangle~,
\label{eq:Z2}
\end{align}
where $\langle I^{(n)}|$ and $|I^{(n)}\rangle$ correspond to the two
irregular punctures on $\mathcal{C}$.

% \begin{figure}
%  \begin{center}
% \begin{tikzpicture}[gauge/.style={circle,draw=black,thick,inner sep=0pt,minimum size=8mm},flavor/.style={rectangle,draw=black,thick,inner sep=0pt,minimum size=8mm},AD/.style={rectangle,draw=black,fill=red!20,thick,inner sep=0pt,minimum size=8mm},auto]

% \node[AD] (1) at (-2,0) {\;$(A_1,D_{2n})$\;};
% \node[gauge] (2) at (0,0) [shape=circle] {\;$2$\;} edge (1);
% \node[AD] (3) at (2,0)  {\;$(A_1,D_{2n})$\;} edge (2);

%   \end{tikzpicture}
% \caption{The quiver diagram of the class $\mathcal{S}$ theory
%   $\mathcal{T}_{\mathcal{C}}$ for $\mathcal{C}$ being sphere with two
%   irregular singularities of rank $n$. Each box stands for
%   $(A_1,D_{2n})$ AD theory whose $SU(2)$ flavor sub-group is gauged.}
% \label{fig:quiver1/2}
%  \end{center}
% \end{figure}

In contrast to the state
corresponding to a regular puncture, the state $|I^{(n)}\rangle$ is not a Virasoro primary
but a linear combination of a primary and its
descendants. This
linear combination is determined to be a
simultaneous solution to a set of equations. There are two
characterizations of this set of equations, and here we follow the
characterization proposed in \cite{Gaiotto:2012sf}:\footnote{The
relation between the characterizations in \cite{Buican:2014hfa} and
\cite{Gaiotto:2012sf} was partially studied in appendix B of \cite{Kanno:2013vi}. }
\begin{align}
L_k|I^{(n)}\rangle = \left\{
\begin{array}{l}
\lambda_k|I^{(n)}\rangle \qquad \text{for} \quad n\leq k\leq 2n
\\[2mm]
\left(\lambda_k + \sum_{\ell = 1}^{n-k}\ell\, c_{\ell+k} \frac{\partial
 }{\partial c_{\ell}}\right)|I^{(n)}\rangle \qquad \text{for} \quad
0\leq k < n
\\
\end{array}
\right.~,
\label{eq:const}
\end{align}
where $\lambda_k$ are constants fixed by $c_0,\cdots,c_n$ %and $c_0\equiv
%\alpha$ 
as
\begin{align}
 \lambda_k = 
\left\{
\begin{array}{l}
-\sum_{\ell=k-n}^n c_\ell c_{k-\ell} \quad \text{for}\quad n<k\leq 2n
\\[2mm]
-\sum_{\ell=0}^{k} c_\ell c_{k-\ell} + (k+1)Qc_k \quad \text{for}\quad k\leq n
\end{array}
\right.~.
\label{eq:lambda}
\end{align}
% \begin{align}
%  \sum_{k=0}^{2n}\frac{\Lambda_k}{z^{k+2}} =
%  -\left(\frac{\alpha}{z}+\sum_{k=1}^n\frac{c_k}{z^{k+1}}\right)^2 - Q
%  \frac{\partial}{\partial z}\left(\frac{\alpha}{z}+\sum_{k=1}^n\frac{c_k}{z^{k+1}}\right)
% \end{align}
% holds as an equality between two rational functions of $z$.
It was
conjectured in \cite{Gaiotto:2012sf} that $|I^{(n)}\rangle$ satisfying the
above equations exists in a highest weight module of Virasoro
algebra.\footnote{Note that the highest weight of this module is not
necessarily equal to $c_0$. Indeed, the highest weight is given by
$\beta_0$ discussed below.}

Since the constraints \eqref{eq:const} are
differential equations, $|I^{(n)}\rangle$ is not completely fixed by
 $c_k$ but depends on the ``boundary
condition'' or the ``asymptotic behavior.'' It was conjectured in \cite{Gaiotto:2012sf} that $|I^{(n)}\rangle$ is
uniquely fixed by specifying  $n$ extra
complex parameters in addition to $c_0,\cdots,c_{n-1}$ and $c_n$. These
extra parameters
characterize the asymptotic behavior of $|I^{(n)}\rangle$ in the small
$c_k$ limit for $k=1,\cdots,n$. We denote these
 extra parameters by $\beta_0,\cdots,\beta_{n-1}$.\footnote{See
 section 3 and appendix B of \cite{Gaiotto:2012sf} for more detail. Some
 explicit examples are also shown in \cite{Nishinaka:2019nuy}.} 
Then the generalized AGT correspondence implies that $c_1,\cdots,c_{n-1}$ are related
to relevant couplings, $\beta_1,\cdots,\beta_{n-1}$ are related to the
VEVs of Coulomb branch operators, and $c_0$ and $\beta_0$ are related to
mass parameters of $(A_1,D_{2n})$ theory. The Liouville charge $Q$ is
identified with $\epsilon_1+\epsilon_2$.

The irregular puncture of integer rank $n$ can be created by colliding $(n+1)$
regular punctures. As a result, the condition \eqref{eq:const} is
obtained in the colliding limit of Virasoro primary operators
\cite{Gaiotto:2012sf}, which we briefly review here for later use.
Let us first consider the state
\begin{align}
	|\phi_n(z_1,\dots,z_n)\rangle \equiv\;\;
 :\!\left(\prod^n_{i=1}V_{\alpha_i}^{L}(z_i)\right)V^L_{\alpha_0}(0)\!:
 |0\rangle~,
\label{eq:phin}
\end{align}
where $V_{\alpha}^{L}(z)$ is the Virasoro primary vertex operator of
conformal weight $\alpha(Q-\alpha)$, and $:\!XY\!:$ is the normal-ordered
product of $X$ and $Y$.
The action of
%stress-energy tensor 
$T_{>}(y)\equiv \sum_{k\geq -1} y^{-k-2}L_k$ on this state is expressed as
%acts this state such 
\begin{align}
\begin{aligned}
	T_{>}(y)|\phi_n(z_1,\dots,z_n)\rangle=\sum_{i=0}^{n}\bigg( \frac{\alpha_i(Q-\alpha_i)}{y-z_i}+\frac{1}{y-z_i}\frac{\partial}{\partial z_i} \bigg)|\phi_n(z_1,\dots,z_n)\rangle~,
\end{aligned}
\label{eq:Tphi}
\end{align}
where $z_0\equiv 0$.
%where $T(y):=\sum_{k\geq-1}y^{-k-2}L_k$,
%  and the generators $L_k\in Vir$ satisfy the commutation relation
% \begin{align}
%  \begin{aligned}
%   &[L_n,L_m]=(n-m)L_{n+m}+\frac{c}{12}(n^3-n)\delta_{n+m}~,\\
%   &[L_n,V_\alpha^L(z)]=(z^{n+1}+(n+1)\alpha(Q-\alpha) z^n)V_\alpha^L(z)~.
%  \end{aligned}
% \end{align}
%
The idea is then to consider a limit $z_i\to 0$ in which the above
action of $T_{>}(y)$ remains well-defined but gives an interesting
result. If we keep $\alpha_i$ finite in the limit, 
$|\phi_n(z_1,\cdots,z_n)\rangle$ just reduces to a single primary vertex
operator acting on the vacuum. A more interesting limit is to take
$z_i\to 0$ and $\alpha_i\to \infty$ with
\begin{align}
 c_k \equiv \sum_{i=0}^n\alpha_i z_i^k \qquad (k=0,\cdots,n)~,
\label{eq:ck}
\end{align}
kept finite. We call this latter limit ``colliding limit.'' It is
straightforward to show that $T_{>}(y)$ acts on the state
\begin{align}
 |I^{(n)}\rangle \equiv \lim_{\text{colliding
 limit}}|\phi_n(z_1,\cdots,z_n)\rangle~,
\label{eq:collided}
\end{align}
as
\begin{align}
 T_{>}(y) |I^{(n)}\rangle =
 \left(\sum_{k=n}^{2n}\frac{\lambda_k}{y^{k+2}} + \sum_{k=0}^{n-1}
 \frac{\lambda_k + \sum_{\ell=1}^{n-k}\ell
 c_{\ell+k}\frac{\partial}{\partial c_\ell}}{y^{k+2}} +
 \frac{L_{-1}}{y}\right)|I^{(n)}\rangle~,
\end{align}
where $\lambda_i$ are defined by \eqref{eq:lambda}. This implies that 
the state  \eqref{eq:collided} satisfies \eqref{eq:const}, and therefore is an
irregular state of rank $n$. Note that, since \eqref{eq:phin} is defined
in terms of the normal-ordered product, \eqref{eq:collided}
gives a well-defined state.

As mentioned already, the irregular state $|I^{(n)}\rangle$ generally
depends on $n$ extra parameters, $\beta_0,\cdots, \beta_{n-1}$,
corresponding to the ``boundary condition'' of a solution to the
differential equations \eqref{eq:const}. These extra parameters
correspond to inserting screening operators in the product
\eqref{eq:phin}. Since screening operators
commute with the Virasoro algebra, this insertion does not break the
conditions \eqref{eq:const}.

\section{Irregular states in $Vir\oplus H$ modules}
\label{sec:V+H}

As reviewed in the previous section, the generalized AGT correspondence
allows us to evaluate the partition function
of $SU(2)$ gauge theory coupled to two $(A_1,D_{2n})$ theories as in
\eqref{eq:Z2}. In this section, we consider an extension of this generalized AGT
correspondence to the case in which the gauge group is $U(2)$ instead of
$SU(2)$. To that end, we start with the $U(2)$ version of the original
AGT correspondence and consider its colliding limit.

\subsection{$U(2)$ version of the original AGT}

The $U(2)$ version of the original AGT correspondence was studied in the
literature. First, it was found in \cite{Alday:2009aq} that the
Nekrasov partition function of a $U(2)$ gauge theory\,
$\mathcal{Z}_{U(2)}$\, is generally related to that of the $SU(2)$ gauge
theory with the same matter content, $\mathcal{Z}_{SU(2)}$, by
\begin{align}
 \mathcal{Z}_{U(2)} = \mathcal{Z}_{SU(2)} \mathcal{Z}_{U(1)}~,
\label{eq:U2}
\end{align}
where $\mathcal{Z}_{U(1)}$ is called ``$U(1)$ factor'' and regarded as
the partition function of the $U(1)$ part of the gauge theory. 
The AGT
correspondence implies that
$\mathcal{Z}_{SU(2)}$ is identical to a conformal block of the 2d
Liouville CFT.

The 2d interpretation of the $U(1)$ factor was then given in
\cite{Alba:2010qc}; $\mathcal{Z}_{U(1)}$ is identical
to a correlation function of chiral vertex operators for an extra
Heisenberg algebra. To be concrete, let us focus on the linear quiver
gauge theory described by the quiver in Fig.~\ref{fig:quiver0}. As
shown in Eq.~\eqref{eq:linear-quiver}, $\mathcal{Z}_{SU(2)}$ is identified with the $(n+2)$-point conformal block of Liouville theory. On
 the other hand, the $U(1)$ factor is identified as
\begin{align}
 \mathcal{Z}_{U(1)} =  \langle V^H_{\alpha_0}(z_0) \cdots
 V^H_{\alpha_{n+1}}(z_{n+1})\rangle ~,
\label{eq:U2AGT}
\end{align}
where $V^H_\alpha(z)\equiv \exp\Big({2(\alpha-Q)
i \sum_{k<0}\frac{a_k}{k}z^{-k}}\Big)\exp\Big({2\alpha
i\sum_{k>0}\frac{a_k}{k}z^{-k}}\Big)$, and the loci $z_k$ of the vertex
operators coincide with those of the Liouville vertex operators in
Eq.~\eqref{eq:linear-quiver}. Here our convention for the Heisenberg
algebra is such that $[a_k,a_\ell] = \frac{k}{2}\delta_{k+\ell,0}$.

Combining \eqref{eq:linear-quiver},\eqref{eq:U2} and \eqref{eq:U2AGT}, we see that
$\mathcal{Z}_{U(2)}$ is identified with the correlator of $(n+2)$ chiral
vertex
operators of the form\footnote{To be precise, one can also insert
screening operators in the Liouville sector here.}
\begin{align}
\widehat{V}_{\alpha}(z)\equiv  V^H_{\alpha}(z) \otimes V^L_{\alpha}(z)~,
\label{eq:VO}
\end{align}
where $V^L_{\alpha}(z)$ is the Virasoro primary vertex operator of
conformal weight $\alpha(Q-\alpha)$ in the
Liouville CFT. 
Thus, the $U(2)$ version of the AGT
correspondence involves the direct sum of Virasoro and
Heisenberg algebras, which we denote by $Vir\oplus H$. Note that $L_k$
and $a_k$ are commutative since we consider the direct sum of the two algebras.
The action of $Vir\oplus H$ on $\widehat{V}_\alpha(z)$
is characterized by
\begin{align}
	[L_n,\widehat{V}_\alpha(z)] &= (z^{n+1}+(n+1)\alpha(Q-\alpha) z^n)\widehat{V}_\alpha(z)~,\\[5pt]
	[a_n,\widehat{V}_\alpha(z)] &= \left\{
  	\begin{array}{l}
   		-i\alpha z^n \widehat{V}_\alpha(z)\quad(n<0)\\
   		i(Q-\alpha)z^n \widehat{V}_\alpha(z)\quad(n>0)
  	\end{array}
  	\right.~.
\end{align}
% It was shown in
% \cite{Alba:2010qc} that highest weight modules of $Vir\times H$
% possess a nice orthogonal basis labeled by a pair of Young diagrams. 

\subsection{$U(2)$ version of the generalized AGT}
\label{subsec:U2gAGT}

We now consider the $U(2)$ version of the generalized AGT
correspondence. Our idea is to start with the $U(2)$ version of the
original AGT, and take the same
colliding limit as the one reviewed in the latter half of  Sec.~\ref{sec:AGT}.

To that end, we start with the quiver gauge theory
described in Fig.~\ref{fig:quiver0} with $U(2)$ gauge groups. The
partition function \eqref{eq:U2} is then identified with the product of
\eqref{eq:linear-quiver} and \eqref{eq:U2AGT}. Note that the loci of the
Heisenberg vertex
operators in \eqref{eq:U2AGT} coincide with those of the Liouville
vertex operators, which reflects the fact that the 4d gauge couplings of
$U(1)\subset U(2)$ and $SU(2)\subset U(2)$ are identical.
 We now take the limit of parameters in which the 4d
theory flows to
the theory described by the quiver in Fig.~\ref{fig:quiver2} with $U(2)$
gauge group. On the 2d side, this corresponds to the colliding limit of
vertex operators \eqref{eq:VO}, and gives rise to an irregular state
$|\widehat{I}^{(n)}\rangle$ of $Vir\oplus H$. The precise definition of
$|\widehat{I}^{(n)}\rangle$ will be given below. The same argument as in
Sec.~\ref{sec:AGT} then leads us to identifying the inner product 
\begin{align}
 \mathcal{Z}^{2\times (A_1,D_{2n})}_{U(2)} = \langle
 \widehat{I}^{(n)}|\widehat{I}^{(n)}\rangle~,
\label{eq:U2gAGT}
\end{align}
as the partition
function of $U(2)$ gauge theory coupled to two $(A_1,D_{2n})$ theories.

Our new irregular state $|\widehat{I}^{(n)}\rangle$ is characterized by
the actions of the Virasoro and Heisenberg algebras on it. These actions can be read off by keeping track of their
actions on $\widehat{V}_{\alpha_k}(z) = V^H_{\alpha_k}(z_k)\otimes
V^L_{\alpha_k}(z_k)$ in the colliding limit.
To see this, 
%let us focus on the current $J(y):=\sum_{k\geq1}y^{-k-1}a_k$ and the
%state
let us consider the state
\begin{align}
 |\widehat{\phi}_n(z_1,\cdots,z_n)\rangle \equiv \;\;
 :\!\left(\prod^n_{i=1}\widehat{V}_{\alpha_i}(z_i)\right)\widehat{V}_{\alpha_0}(0)\!:
 |0\rangle~.
\label{eq:tilde-phin}
\end{align}
The action of $T_{>}(y)$ on this state is the same as in Eq.~\eqref{eq:Tphi}.
Similarly, the action of $J_{>}(y)\equiv \sum_{k\geq 1}y^{-k-1}a_k$ is
written as
\begin{align}
 J_{>}(y)|\widehat{\phi}_n(z_1,\cdots,z_n)\rangle  = -\sum_{i=0}^n
 \frac{i(Q-\alpha_i)z_i}{y(y-z_i)}|\widehat{\phi}(z_1,\cdots,z_n)\rangle~,
\end{align}
where $z_0\equiv 0$.
We now take the colliding limit $z_i\to 0$ and $\alpha_i\to \infty$ with
\eqref{eq:ck} kept fixed. The irregular state
$|\widehat{I}^{(n)}\rangle$ is now defined by
\begin{align}
 |\widehat{I}^{(n)}\rangle \equiv \lim_{\text{colliding limit}}|\widehat{\phi}_n(z_1,\cdots,z_n)\rangle~.
\end{align}
It is straightforward to show that $T_{>}(y)$ and $J_{>}(y)$ act on this
state as
% \begin{align}
% \begin{aligned}
% 	|\widehat{\phi}_n(z_1,\dots,z_n)\rangle=\widehat{C}_n(z_1,\dots,z_n)\lim_{z_0\rightarrow 0}\prod^n_{i=0}V_{\alpha_i}(z_i)|0\rangle~,
% \end{aligned}
% \end{align}
% where $\widehat{C}_n(z_1,\dots,z_n)$ is also the normalization factor.
%  Here the generators $L_n\in\text{Vir}$ and $a_n\in\mathcal{H}$ are commutative, and the commutation relations are obtained as
% \begin{align}
% \begin{aligned}
% 	&[L_n,V_\alpha(z)]=(z^{n+1}+(n+1)\alpha(Q-\alpha) z^n)V_\alpha(z)~,\\[5pt]
% 	&[a_n,V_\alpha(z)]=\left\{
%   	\begin{array}{l}
%    		-i\alpha z^n V_\alpha(z)\quad(n<0)\\
%    		i(Q-\alpha)z^nV_\alpha(z)\quad(n>0)
%   	\end{array}
%   	\right.~.
% \end{aligned}
% \end{align}
\begin{align}
 T_{>}(y)|\widehat{I}^{(n)}\rangle &= \left(\sum_{k=n}^{2n}
 \frac{\lambda_k}{y^{k+2}}+
 \sum_{k=0}^{n-1}\frac{\lambda_k+\sum_{\ell=1}^{n-k}\ell
 c_{\ell+k}\frac{\partial}{\partial c_\ell}}{y^{k+2}} +
 \frac{L_{-1}}{y}\right)~,
\\
J_{>}(y)|\widehat{I}^{(n)}\rangle &= \sum_{k=1}^{n}\frac{-ic_k}{y^{k+1}}|\widehat{I}^{(n)}\rangle~,
\end{align}
% In the colliding limit, the current acting on $|\widehat{\phi}_n(z_1,\dots,z_n)\rangle$ reduces to
% \begin{align}
% \begin{aligned}
% 	J(y)|\widehat{\phi}_n(z_1,\dots,z_n)\rangle= \sum_{i=0}^n \frac{i(Q-\alpha_i) z_i}{y(y-z_i)} |\widehat{\phi}_n(z_1,\dots,z_n)\rangle\rightarrow\sum_{k=1}^{n}\frac{-ic_k}{y^{k+1}}|\widehat{\phi}_n(z_1,\dots,z_n)\rangle~,
% \end{aligned}
% \end{align}
where $\lambda_k$ and $c_k$ are given by %(\ref{eq:calpha}).
\eqref{eq:lambda} and \eqref{eq:ck}, respectively. 
% Thus we obtain the constraints of $|\widehat{I}^{(n)}\rangle$
% \begin{align}
% 	J(y)|\widehat{I}^{(n)}\rangle=\sum_{k=1}^{n}\frac{-ic_k}{y^{k+1}}|\widehat{I}^{(n)}\rangle~.
% \end{align}
% Since $L_n$ commutes to $a_n$, the stress-energy tensor further leads to the similar constraints
% \begin{align}
% \begin{aligned}
% 	T(y)|\widehat{I}^{(n)}\rangle=\Bigg( \sum_{j=n}^{2n}\frac{\lambda_j}{y^{j+2}}+\sum_{k=0}^{n-1}\frac{\lambda_k+\sum_{\ell=1}^{n-k}\ell c_{\ell+k}\frac{\partial}{\partial c_{\ell}}}{y^{k+2}}+\frac{L_{-1}}{y} \Bigg)|\widehat{I}^{(n)}\rangle~.
% \end{aligned}
% \end{align}
% %\noindent
% %
% %
From the above result, we see that $Vir\oplus H$ acts on
$|\widehat{I}^{(n)}\rangle$ as
\begin{align}
L_k|\widehat{I}^{(n)}\rangle &= \left\{
\begin{array}{l}
\lambda_k|\widehat{I}^{(n)}\rangle \qquad \text{for} \quad n\leq k\leq 2n
\\[2mm]
\left(\lambda_k + \sum_{\ell = 1}^{n-k}\ell\, c_{\ell+k} \frac{\partial
 }{\partial c_{\ell}}\right)|\widehat{I}^{(n)}\rangle \qquad \text{for} \quad
0\leq k < n
\\
\end{array}
\right.~,
\label{eq:const2}
\\[3mm]
 a_k |\widehat{I}^{(n)}\rangle &=
\left\{
\begin{array}{l}
\;\; \;\;0 \qquad \text{for}\quad n<k
\\[2mm]
 -ic_k|\widehat{I}^{(n)}\rangle \qquad \text{for}\quad
 1\leq k\leq n
 \\
\end{array}
\right.~.
\label{eq:const3}
\end{align}
%where $\lambda_k$ are given by \eqref{eq:lambda}.

Note here that the above characterization of the irregular state does
not fix the overall normalization, as in the case of the $SU(2)$-version
of the generalized AGT correspondence. This means an ambiguity in
the computation of
the perturbative part of the partition function
\eqref{eq:U2gAGT}. However, it turns out that the instanton part of the partition
function can be unambiguously computed, as will be discussed in the following sections.

Note also that $|\widehat{I}^{(n)}\rangle$ is by definition decomposed into
the Virasoro part and the Heisenberg part as $|\hat{I}^{(n)}\rangle =
|I^{(n)}\rangle\otimes |I^{(n)}_H\rangle$, where  $|I^{(n)}\rangle$ is
the irregular state of Virasoro algebra that was reviewed in
Sec.~\ref{sec:AGT}. While $|I^{(n)}\rangle$ depends on $n$ extra
parameters in addition to $c_0,\cdots,c_{n-1}$ and $c_n$,
%\cite{Gaiotto:2012sf} 
 the state $|I^{(n)}_H\rangle$ is uniquely fixed by
$c_1,\cdots,c_{n-1}$ and $c_n$ up to a prefactor. We here write down its explicit
expression:
\begin{align}
 |I^{(n)}_H\rangle = \exp\left(-2i\sum_{k=1}^n \frac{c_k}{k} a_{-k}
 \right)|0\rangle~,
\label{eq:IH}
\end{align}
where 
%$|\text{h.w.}\rangle_H$ is a highest weight state such that
%$a_k|\text{h.w.}\rangle_H = 0$ for $k>0$, and 
the prefactor is fixed so that $|I^{(n)}_H\rangle$ reduces to
$|0\rangle$ when $c_k=0$. Since $|I^{(n)}_H\rangle$ and
$|I^{(n)}\rangle$ are respectively in a highest weight module of $Vir$
and $H$, we see that $|\widehat{I}^{(n)}\rangle$ is a state in a highest
weight module of $Vir\oplus H$. Note that
the Heisenberg sector has no possible insertion of screening operators, and therefore
\eqref{eq:IH} is the unique expression for  $|I^{(n)}_H\rangle$ up to a prefactor. Indeed, the constraints
\eqref{eq:const3} are eigenstate equations, whose solution is fixed (up
to the normalization) by
$\{c_k\}$ without
specifying a ``boundary condition.''

Given the $U(2)$-version of the generalized AGT correspondence
\eqref{eq:U2gAGT}, one can now study the decomposition of
$\mathcal{Z}_{U(2)}^{2\times (A_1,D_{2n})}$ as a sum over pairs of Young
diagrams as in Eq.~\eqref{eq:Nek2}. In the next section, we explicitly
evaluate this decomposition to read off the factor
$\mathcal{Z}_{Y_1,Y_2}^{(A_1,D_{2n})}$ in \eqref{eq:Nek2}.

\section{Nekrasov-type formula for AD matter}
\label{sec:Nek-AD}

Here we read off the factor
$\mathcal{Z}_{Y_1,Y_2}^{(A_1,D_{2n})}$ in \eqref{eq:Nek2} from
the $U(2)$-version of the generalized AGT correspondence
\eqref{eq:U2gAGT}. This factor can be interpreted as the contribution of
the $(A_1,D_{2n})$ theory at the fixed point corresponding to
$(Y_1,Y_2)$ on the $U(2)$ instanton moduli space.

\subsection{Decomposition}

To read off $\mathcal{Z}_{Y_1,Y_2}^{(A_1,D_{2n})}$, we first decompose \eqref{eq:U2gAGT} as a sum over
pairs of Young diagrams, using a nice basis of highest weight
modules of $Vir\oplus H$ that was found in
\cite{Alba:2010qc}. To describe it, let us denote by 
\begin{align}
 \mathcal{Z}_{Y_1,Y_2;W_1,W_2}^\text{bifund}(a,b,\alpha)~,
\label{eq:bifund0}
\end{align}
the contribution to the Nekrasov partition function from a
bi-fundamental hypermultiplet of $U(2)\times U(2)$. Here, $(a,b)$
stands for the VEVs of Coulomb branch operators in the vector multiplet
for $SU(2)\times SU(2) \subset U(2)\times U(2)$, and $\alpha$ is a mass parameter. The explicit
expression for \eqref{eq:bifund0} is written in Appendix
\ref{app:Nek}. 
 It was shown in \cite{Alba:2010qc} that there
exists an orthogonal basis, $|a;Y_1,Y_2\rangle$, of the highest weight module of $Vir \oplus H$ such that
\begin{align}
\frac{\langle a;Y_1,Y_2| V_\alpha(1) |b;W_1,W_2\rangle}{\langle
 a|V_\alpha(1)|b\rangle} &=
 \mathcal{Z}_{Y_1,Y_2;W_1,W_2}^\text{bifund}(a,b,\alpha)~,
\label{eq:basis}
\end{align}
where  $|a\rangle$ is the highest weight state satisfying $L_0|a\rangle =
\Delta(a)|a\rangle$ and $L_n|a\rangle = a_n|a\rangle = 0$ for $n>0$,
$V_\alpha(z)$ is the vertex operator shown in \eqref{eq:VO}, and $Y_k$
and $W_k$ are arbitrary Young diagrams.
Note that the conjugate $\langle a;Y_1,Y_2|$ is not the usual
hermitian conjugate of $|a;Y_1,Y_2\rangle$; it is obtained
by expanding $|a;Y_1,Y_2\rangle$ as a linear combination of $L_{-k_1}^{m_1}\cdots L_{-k_p}^{m_p}
a_{-\ell_1}^{n_1}\cdots a_{-\ell_q}^{n_q}|a\rangle$, and then replacing each such
state with $\langle a| L_{k_p}^{m_p}\cdots L_{k_1}^{m_1}
a_{\ell_q}^{n_q}\cdots a_{\ell_1}^{n_1}$ without changing the coefficients.
 The first few examples of
$|a;Y_1,Y_2\rangle$ are presented in Appendix \ref{app:basis}. 
It was also
shown in \cite{Alba:2010qc} that $|a;Y_1,Y_2\rangle$ is generally a
linear combination of descendants of $|a\rangle$ at level
$(|Y_1|+|Y_2|)$.\footnote{Here, the ``level'' is defined by the sum of
the levels of the Virasoro and Heisenberg parts. For example, the level
of $L_{-k_1}(L_{-k_2})^3a_{-\ell}|a\rangle$ is $k_1+3k_2+\ell$.}

As discussed in \cite{Alba:2010qc},
%  and also reviewed in Appendix
% \ref{app:basis},
 the condition \eqref{eq:basis} and the fact that
$\mathcal{Z}_{Y_1,Y_2}^\text{vec}(a) =
1/\mathcal{Z}_{Y_1,Y_2;Y_1,Y_2}^\text{bifund}(a,a,0)$ imply
\begin{align}
 {\bf 1} =
 \sum_{Y_1,Y_2}\mathcal{Z}_{Y_1,Y_2}^\text{vec}(a)\,|a;Y_1,Y_2\rangle\langle
 a;Y_1,Y_2|~,
\label{eq:dec-of-1}
\end{align}
on the highest weight $Vir\oplus H$-module associated with
$|a\rangle$. Note again that $\langle a;Y_1,Y_2|$ is not the usual
conjugate of $|a;Y_1,Y_2\rangle$.
Let us now take $|a\rangle$ to be the highest weight state of the $Vir\oplus H$-module that
includes $|\widehat{I}^{(n)}\rangle$. Then, by inserting \eqref{eq:dec-of-1}
in Eq.~\eqref{eq:U2gAGT}, one obtains the
following decomposition:
\begin{align}
 \mathcal{Z}_{U(2)}^{2\times (A_1,D_{2n})} =
 \sum_{Y_1,Y_2}\mathcal{Z}_{Y_1,Y_2}^\text{vec}(a)\,\langle \widehat{I}^{(n)}|a;Y_1,Y_2\rangle\langle
 a;Y_1,Y_2|\widehat{I}^{(n)}\rangle~.
\label{eq:decomp1}
\end{align}

We interpret the above expression as a sum over fixed points on
the moduli space of $U(2)$ instantons, and $\langle
a;Y_1,Y_2|\widehat{I}^{(n)}\rangle$ and $\langle
\hat{I}^{(n)}|a;Y_1,Y_2\rangle$ as the contributions of the
$(A_1,D_{2n})$ theories corresponding to $|\hat{I}^{(n)}\rangle$
and $\langle \hat{I}^{(n)}|$, respectively. Note that this
particularly implies that $\langle a;\emptyset,\emptyset|\hat{I}^{(n)}\rangle$ and $\langle
\hat{I}^{(n)}|a;\emptyset,\emptyset\rangle$ are the partition function of $(A_1,D_{2n})$ theory
with its flavor symmetry un-gauged. Since $|a;\emptyset,\emptyset\rangle
= |a\rangle$ \cite{Alba:2010qc}, this is indeed consistent with
\eqref{eq:ZAD1}. 
% Our interpretation
% above can be regarded as its generalization to the case in which the flavor symmetry is
% gauged by the $U(2)$ vector multiplet.

\subsection{Identification of $\mathcal{Z}^{(A_1,D_{2n})}_{Y_1,Y_2}$}
\label{subsec:ZAD}

Comparing \eqref{eq:decomp1} with \eqref{eq:Nek2}, we see that it is natural to interpret
\begin{align}
 \mathcal{Z}^{(A_1,D_{2n})}_{Y_1,Y_2} \sim \frac{\langle
a;Y_1,Y_2|\hat{I}^{(n)}\rangle}{\langle a|\hat{I}^{(n)}\rangle}~,\qquad
 \tilde{\mathcal{Z}}^{(A_1,D_{2n})}_{Y_1,Y_2} \sim \frac{\langle
\hat{I}^{(n)}|a;Y_1,Y_2\rangle}{\langle \hat{I}^{(n)}|a\rangle}~,
\label{eq:proportionality}
\end{align}
with possible proportionality constants. Note that the denominators in
\eqref{eq:proportionality} are necessary for
$\mathcal{Z}^{(A_1,D_{2n})}_{\emptyset,\emptyset} =
\tilde{\mathcal{Z}}_{\emptyset,\emptyset}^{(A_1,D_{2n})} = 1$.
 
Here, $|\widehat{I}^{(n)}\rangle$ and $\langle \widehat{I}^{(n)}|$
correspond to two different $(A_1,D_{2n})$ theories. Indeed, as seen
from their colliding-limit derivation, these $(A_1,D_{2n})$ theories
are differently coupled to the $U(1)$-part of the gauge
group. Therefore, $\mathcal{Z}_{Y_1,Y_2}^{(A_1,D_{2n})}$ and
$\tilde{Z}^{(A_1,D_{2n})}_{Y_1,Y_2}$ are not identical. 

% To that end, let us
% first identify the relation between the 2d and 4d parameters. Recall
% that $|\hat{I}^{(n)}\rangle$ depends on $c_1,\cdots, c_n$ and $c_0\equiv
% \alpha$ together with
% $n$ extra parameters. One of the extra parameters fixes the highesttj
% weight, $a$, of the $Vir\times H$-module that contains
% $|\hat{I}^{(n)}\rangle$, and therefore $\langle a;Y_1,Y_2|\hat{I}^{(n)}\rangle$
% is completely fixed by these $(2n+1)$ parameters. 

In the rest of this sub-section, we make the relations \eqref{eq:proportionality} more
precise. 
To that end, we first focus on the left relation, and read off how the
4d and 2d parameters are related. Note that
 $\langle
 a;Y_1,Y_2|\hat{I}^{(n)}\rangle$ on the RHS depends on $(2n+1)$ parameters. Indeed,
 $|\hat{I}^{(n)}\rangle$ depends on $n$ extra parameters
 $\beta_0,\cdots,\beta_{n-1}$ in addition to
 $c_0,\cdots,c_{n}$, as reviewed in
 Sec.~\ref{subsec:U2gAGT}. One of these extra parameters fixes the highest weight, $a$, of the
 $Vir\oplus H$-module that includes $|\hat{I}^{(n)}\rangle$, and therefore
 $\langle a;Y_1,Y_2|\hat{I}^{(n)}\rangle$ is completely fixed by these $(2n+1)$
 parameters. On the other hand,
 $\mathcal{Z}^{(A_1,D_{2n})}_{Y_1,Y_2}(a,m,\pmb{d},\pmb{u})$ on the LHS depends only
on $2n$ parameters, $a,m,\pmb{d}=(d_1,\cdots,d_{n-1})$ and
$\pmb{u}=(u_1,\cdots, u_{n-1})$. Here, $a$ and $\pmb{u}$ are the VEVs of
Coulomb branch operators, $\pmb{d}$ are relevant couplings and $m$ is
a mass parameter. Therefore, there is a discrepancy in the number of
 parameters between the 2d and 4d sides.

To see this discrepancy more explicitly, let us identify the precise relation between the 2d and 4d parameters. We start with the Seiberg-Witten
(SW) curve of the $(A_1,D_{2n})$
theory \cite{Bonelli:2011aa, Xie:2012hs}
\begin{align}
 x^2 = \frac{a^2}{z^2} + \sum_{k=1}^{n-1}\frac{u_k}{z^{n+2-k}} +
 \frac{m}{z^{n+2}} + \sum_{k=1}^{n-1}\frac{d_k}{z^{2n+2-k}} +
 \frac{1}{z^{2n+2}}~,
\label{eq:curve1}
\end{align}
where the SW 1-form is given by $xdz$.\footnote{Here $a$ is regarded as the mass parameter corresponding to a
flavor $SU(2)$ sub-group of the $(A_1,D_{2n})$ theory.} This curve is
identified, on the 2d side, as the classical limit $\epsilon_i\to 0$ of the following
\cite{Alday:2009aq}
\begin{align}
 x^2 = -\frac{\langle a |T(z)|\widehat{I}^{(n)}\rangle}{\langle
 a|\widehat{I}^{(n)}\rangle} = -\frac{\Delta (a)}{z^2} + \cdots+ \frac{2c_nc_{n-1}}{z^{2n+1}}
 +\frac{c_n^2}{z^{2n+2}}~,
\label{eq:curve2}
\end{align}
up to a change of variables that preserves the SW 1-form. Note that the
RHS of the above equation can be explicitly evaluated via
Eq.~\eqref{eq:const2}.\footnote{Recall here that, to recover the full
$\epsilon_i$-dependence, one needs to perform a replacement
corresponding to \eqref{eq:recover}. For the irregular state
$|I^{(n)}\rangle$, this replacement implies $c_k\to
c_k/\sqrt{\epsilon_1\epsilon_2}$ as seen
from its colliding-limit derivation.}
We here change the
variables in \eqref{eq:curve2} as $z\to (c_n)^{\frac{1}{n}}\,z$ and $x\to (c_n)^{-\frac{1}{n}}\,x$ so that the
coefficient of $1/z^{2n+2}$ is $1$. Comparing the classical limit of the
resulting equation
with \eqref{eq:curve1}, we obtain the relation between the 2d and 4d
parameters.\footnote{Note here that $\Delta(a)$ reduces to $-a^2$ in the classical limit
$\epsilon_1,\epsilon_2\to 0$.} To make this relation simple, let us define on the 2d side
\begin{align}
 \gamma_k\equiv \frac{c_k}{(c_n)^{\frac{k}{n}}}~,
\end{align}
for $k=0,\cdots,n-1$. In terms of these variables, the relation between
the 2d and 4d parameters is expressed as
\begin{align}
 d_k 
&=
\sum_{\ell=n-k}^n \gamma_\ell \gamma_{2n-k-\ell}~,\quad m =
 \sum_{\ell= 0}^n\gamma_\ell\gamma_{n-\ell}~,\quad u_k = \sum_{\ell =0}^{n-k} \gamma_\ell \gamma_{n-k-\ell} - \sum_{\ell=1}^{k}\ell \gamma_{\ell +
 n-k}\frac{\partial}{\partial \gamma_\ell}\mathcal{F}_{(A_1,D_{2n})}~,
%+ (n+1)Q
% &=
% -\sum_{\ell=n-k}^n \frac{c_\ell c_{2n-k-\ell}}{(c_n)^{\frac{2n-k}{n}}}~,\qquad m =
%  -\sum_{\ell= 0}^n\frac{c_\ell c_{n-\ell}}{c_n}- (n+1)Q~,
% \\
% u_k &= -\sum_{\ell =0}^{n-k} \Big(\gamma_\ell \gamma_{n-k-\ell} + (n-k+1)Q\gamma_{n-k}\Big) + \sum_{\ell=1}^{k}\ell \gamma_{\ell +
%  n-k}\frac{\partial}{\partial \gamma_\ell}\log \langle
%  a|\hat{I}^{(n)}\rangle~,
% &= -\sum_{\ell =0}^{n-k} \frac{c_\ell c_{n-k-\ell} +
%  (n-k+1)Qc_{n-k}}{(c_n)^{\frac{n-k}{n}}} + \sum_{\ell=1}^{k}\ell \frac{c_{\ell +
%  n-k}}{(c_n)^{\frac{n-k}{n}}}\frac{\partial}{\partial c_\ell}\log \langle
%  a|\hat{I}^{(n)}\rangle~,
\label{eq:coupling}
\end{align}
where the derivatives
$\partial/\partial \gamma_\ell$ are defined with
$\vec{\gamma}\equiv (\gamma_0,\cdots,\gamma_{n-1})$ and $c_n$ taken
as independent variables, and $\mathcal{F}_{(A_1,D_{2n})}$ is the
classical limit of $\log \langle a|\hat{I}^{(n)}\rangle$.
These expressions imply that, when one takes
$\vec{\gamma}$ and $c_n$ as independent variables, all the 4d
parameters are independent of $c_n$.\footnote{To prove this statement,
one needs to show that $\frac{\partial}{\partial \gamma_\ell}\mathcal{F}_{(A_1,D_{2n})}$ is independent
of $c_n$. This can be shown as follows. As we will show below, $\langle
a;Y_1,Y_2|\widehat{I}^{(n)}\rangle = 
(c_n)^{\frac{\Delta_a-\Delta_{c_0}+|Y_1|+|Y_2|}{n}}
f_{Y_1,Y_2}(\vec{\gamma})$ for a function $f_{Y_1,Y_2}(\vec{\gamma})$ independent of $c_n$. Setting $Y_1=Y_2=\emptyset$, we find
$\log \langle a|\widehat{I}^{(n)}\rangle =
\frac{\Delta_a-\Delta_{c_0}}{n}\log c_n  + \log
f_{\emptyset,\emptyset}(\vec{\gamma})$. This implies that
$\frac{\partial}{\partial \gamma_\ell}\log\langle a|\widehat{I}^{(n)}\rangle=\frac{\partial}{\partial \gamma_\ell}\log
f_{\emptyset,\emptyset}(\vec{\gamma})$. Since this is independent of
$c_n$, its classical limit $\frac{\partial}{\partial
\gamma_\ell}\mathcal{F}_{(A_1,D_{2n})}$ is also independent of $c_n$ when
written in terms of $\vec{\gamma}$ and $c_n$.}
% \footnote{To prove
% this statement, one needs to show that $\frac{\partial}{\partial \gamma_k}\log\langle a|\hat{I}^{(n)}\rangle$ is
% idependent of $c_n$ when $\gamma_1,\cdots,\gamma_{n-1}$ and $c_n$ are
% taken as independent variables. This can be shown by noting that, when considering
% $|\hat{I}^{(n)}\rangle$ modulo its normalization, the rescale \eqref{eq:rescale} is
% equivalent to $L_k \to \zeta^{-k}L_k$ and $a_k\to \zeta^{-k}a_k$, as
% seen from \eqref{eq:const2} and \eqref{eq:const3}. It is clear that this latter set of
% transformations preserves
% $\log\langle a|\hat{I}^{(n)}\rangle$. The normalization of
% $|\hat{I}^(n)\rangle$ might change under \eqref{eq:rescale}, but that
% only gives rise to a constant shift of $\log \langle
% a|\hat{I}^{(n)}\rangle$. }
  This reflects the conformal invariance of $(A_1,D_{2n})$, and explains the discrepancy in
the number of parameters between the 2d and 4d sides. 

The above discussion implies that, for the left relation in
\eqref{eq:proportionality} to be an equality, the $c_n$-dependence of
the RHS needs to be canceled by a constant of proportionality.
To identify this proportionality constant, let us consider
\begin{align}
n\,c_n\left.\frac{\partial}{\partial c_n}\right|_{\vec{\gamma}}\langle
 a;Y_1,Y_2|\hat{I}^{(n)}\rangle~,
\label{eq:cdc1}
\end{align}
where $\partial/\partial c_n|_{\vec{\gamma}}$ is the derivative with respect to
$c_n$ with $\vec{\gamma} = (\gamma_0,\cdots,\gamma_{n-1})$ kept
fixed. From \eqref{eq:const2}, we see that this is identical to
\begin{align}
 \langle a;Y_1,Y_2|(L_0 -\Delta_{c_0}) |\widehat{I}^{(n)}\rangle =
 \left(\Delta_a -\Delta_{c_0} +|Y_1|+|Y_2|\right)\langle a;Y_1,Y_2|\hat{I}^{(n)}\rangle~.
\label{eq:cdc2}
\end{align}
The equality between the above two implies that $\langle
a;Y_1,Y_2|\hat{I}^{(n)}\rangle  \sim (c_n)^{\frac{\Delta_a-\Delta_{c_0}
+  |Y_1|+|Y_2|}{n}}$, and
therefore the ratio $(c_n)^{-\frac{|Y_1|+|Y_2|}{n}}\langle
a;Y_1,Y_2|\hat{I}^{(n)}\rangle/\langle a|\hat{I}^{(n)}\rangle$ is independent of $c_n$ when written in
terms of
$\vec{\gamma}$ and $c_n$.\footnote{Recall here that $\langle
a;\emptyset,\emptyset| = \langle a|$.} This
suggests the following identification
\begin{align}
 \mathcal{Z}^{(A_1,D_{2n})}_{Y_1,Y_2} =
  (\zeta c_n)^{-\frac{|Y_1|+|Y_2|}{n}}\frac{\langle
a;Y_1,Y_2|\hat{I}^{(n)}\rangle}{\langle a|\hat{I}^{(n)}\rangle}~,
\label{eq:proposal1}
\end{align}
where $\zeta$ is a possible numerical constant independent of all
variables. Note that $\zeta$ above can be absorbed by rescaling the
instanton factor $\Lambda$.
% up to a possible numerical factor that can be absorbed by rescaling the
% instanton factor $\Lambda$.
%  In the rest of this
% paper,
% we sometimes set $\zeta=-1/2$ to avoid numerical factors.

A parallel discussion shows that the parameters of
$\tilde{\mathcal{Z}}_{Y_1,Y_2}(a,\tilde{m},\tilde{\pmb{d}},\tilde{\pmb{u}})$
are related to those of $\langle \hat{I}^{(n)}|a;Y_1,Y_2\rangle$ by
a similar relation to \eqref{eq:coupling}. As $\langle
a;Y_1,Y_2|\hat{I}^{(n)}\rangle$ depends on $c_0,\cdots,c_n$ and
$\beta_0,\cdots,\beta_{n-1}$, $\langle \hat{I}^{(n)}|a;Y_1,Y_2\rangle$
also depends on $(2n+1)$ parameters, which we denote by
$\tilde{c}_0,\cdots,\tilde{c}_{n}$ and
$\tilde{\beta}_0,\cdots,\tilde{\beta}_{n-1}$. 
% Then $\langle
% \hat{I}^{(n)}|a;Y_1,Y_2\rangle$ depends on
% $\tilde{c}_0^{\,*},\cdots,\tilde{c}_{n}^{\,*}$ together with $n$ extra
% parameters.
 From the same argument as above, we see that
\begin{align}
\tilde{\mathcal{Z}}^{(A_1,D_{2n})}_{Y_1,Y_2} =
 (-\zeta\tilde{c}_n^{\,*})^{-\frac{|Y_1|+|Y_2|}{n}}\frac{\langle
 \hat{I}^{(n)}|a;Y_1,Y_2\rangle}{\langle \hat{I}^{(n)}|a\rangle}~,
\label{eq:proposal2}
\end{align}
where $\tilde{c}_n^{\,*}$ is the complex conjugate of
$\tilde{c}_n$.\footnote{The opposite sign in the bracket in $(-\tilde{c}_n^*)^{-\frac{|Y_1|+|Y_2|}{n}}$ can be
understood as follows. First recall that $\langle a;Y_1,Y_2|$ in
\eqref{eq:proposal2} is {\it not} the usual conjugate of
$|a;Y_1,Y_2\rangle$, as discussed in \cite{Alba:2010qc}. Indeed,
$\langle a;Y_1,Y_2|$ is obtained by expanding $|a;Y_1,Y_2\rangle$ as a
linear combination of $L_{-k_1}^{m_1}\cdots
L_{-k_p}^{m_p}a_{-\ell_1}^{n_1}\cdots a_{-\ell_q}^{n_q}|a\rangle$, and
then replacing each such state with $\langle a|L_{k_p}^{m_p}\cdots
L_{k_1}^{m_1}a_{\ell_q}^{n_q}\cdots a_{\ell_1}^{n_1}$ without changing
the coefficients. This implies that, $\langle
\hat{I}^{(n)}|a;Y_1,Y_2\rangle$ is obtained from $\langle
a;Y_1,Y_2|\hat{I}^{(n)}\rangle$ by replacing
$\langle a|L_{k_p}^{m_p}\cdots
L_{k_1}^{m_1}a_{\ell_q}^{n_q}\cdots
a_{\ell_1}^{n_1}|\hat{I}^{(n)}\rangle
$ with $\langle \hat{I}^{(n)}|L_{-k_p}^{m_p}\cdots
L_{-k_1}^{m_1}a_{-\ell_q}^{n_q}\cdots a_{-\ell_1}^{n_1}|a\rangle$. From \eqref{eq:const2} and \eqref{eq:const3}, we see that this is
equivalent to the replacement
\begin{align}
 c_k \longrightarrow -\tilde{c}_k^*~,\qquad Q\to -Q^*
\end{align}
for $k=0,\cdots,n$. In particular, $c_n$ in \eqref{eq:proposal1} is
replaced by $-\tilde{c}_n^*$.}

In the above identifications, $\mathcal{Z}^{(A_1,D_{2n})}_{Y_1,Y_2}$ and
$\tilde{\mathcal{Z}}^{(A_1,D_{2n})}_{Y_1,Y_2}$ are regarded as the contribution
of the $(A_1,D_{2n})$ theory corresponding to $|\widehat{I}^{(n)}\rangle$
and $\langle \widehat{I}^{(n)}|$, respectively. As discussed at the beginning, these two $(A_1,D_{2n})$ theories have different couplings to the $U(1)$
part of the gauge group.
As we will see in Sec.~\ref{subsec:check}, the difference between $\mathcal{Z}^{(A_1,D_{2n})}_{Y_1,Y_2}$ and
$\tilde{\mathcal{Z}}^{(A_1,D_{2n})}_{Y_1,Y_2}$ is an AD counterpart of the difference between the fundamental and
anti-fundamental hypermultiplets of $U(2)$.

% Note also that \eqref{eq:proposal1} and \eqref{eq:proposal2} contain the
% ``perturbative'' part arising from the $(A_1,D_{2n})$ theory. Indeed,
% these factors for $(Y_1,Y_2) =(\emptyset,\emptyset)$ are the partition
% function of the $(A_1,D_{2n})$ theory with its flavor symmetry
% un-gauged, and therefore regarded as the classical part. To extract the
% instanton part, one can take the ratios
% $\mathcal{Z}_{Y_1,Y_2}^{(A_1,D_{2n})}/\mathcal{Z}_{\emptyset,\emptyset}^{(A_1,D_{2n})}$
% and $\widetilde{\mathcal{Z}}_{Y_1,Y_2}^{(A_1,D_{2n})}/\widetilde{\mathcal{Z}}_{\emptyset,\emptyset}^{(A_1,D_{2n})}$.

% In the next sub-section, we show that our proposals \eqref{eq:proposal1}
% and \eqref{eq:proposal2} reproduce the correct $\Lambda$-dependence of
% \eqref{eq:Nek2}. 

% \vskip.5cm
% \note{Takahiro}{Here we should write how to fix $c_n$. Indeed, we are
% setting $c_n=-1/2$ in our code, but we are not actually sure what the most
% natural value is... Well, it is true that setting $c_n=-1/2$ implies
% $\Lambda = 1$ in the case of $n=1$, but we are not sure if that is the
% case for $n\geq 2$ as well.}

\subsection{Identification of $\Lambda$}

The identifications \eqref{eq:proposal1} and \eqref{eq:proposal2} imply that
\eqref{eq:decomp1} is re-expressed as
\begin{align}
 \mathcal{Z}^{2\times (A_1,D_{2n})}_{Y_1,Y_2} &=
 \mathcal{Z}_\text{pert}\sum_{Y_1,Y_2}(-\zeta^2 c_n\tilde{c}_n^{\,*})^{\frac{|Y_1|+|Y_2|}{n}}\mathcal{Z}^\text{vec}_{Y_1,Y_2}(a)\mathcal{Z}_{Y_1,Y_2}^{(A_1,D_{2n})}(a,m,\pmb{d},\pmb{u})\tilde{\mathcal{Z}}^{(A_1,D_{2n})}_{Y_1,Y_2}(a,\tilde{m},\tilde{\pmb{d}},\tilde{\pmb{u}})~,
\label{eq:decomp}
\end{align}
with the perturbative part $\mathcal{Z}_\text{pert}\equiv \langle \hat{I}^{(n)}|a\rangle\langle a|\hat{I}^{(n)}\rangle$.
Comparing \eqref{eq:decomp} with \eqref{eq:Nek2}, we identify the
4d dynamical scale as
\begin{align}
 \Lambda^2 = - \zeta^2 c_n\tilde{c}_n^{\,*}~.
\label{eq:Lambda}
\end{align} 
Recall that the $(A_1,D_{2n})$ sector is independent of $c_n$ (and $\tilde{c}_n^{\,*}$) as a
result of its conformal invariance. Here, the $U(2)$ gauge coupling
breaks this conformal invariance through the dynamical scale, and
therefore it is natural that $\Lambda$ depends on $c_n$ and
$\tilde{c}_n^{\,*}$.

The identification \eqref{eq:Lambda} is also consistent with the SW
curve. The curve of the theory shown in Fig.~\ref{fig:quiver2} is written as
\cite{Bonelli:2011aa}
\begin{align}
 x^2 = \Lambda_0^2 z^{2n-2} + \cdots + \frac{\Lambda_0^2}{z^{2n+2}}~,
\label{eq:curve3}
\end{align}
where $\Lambda_0$ is a dynamical scale that can differ from
$\Lambda$ by a numerical factor, and the ellipsis stands for a Laurent polynomial of $z$ which is less singular
than $z^{2n-2}$ at $z=\infty$ and than $1/z^{2n+2}$ at
$z=0$.\footnote{The SW 1-form is again given by $xdz$.} Now,
by the same argument as around Eq.~\eqref{eq:curve2}, this curve is
identified as
\begin{align}
 x^2 = -\frac{\langle \hat{I}^{(n)}| T(z)|\hat{I}^{(n)}\rangle}{\langle
 \hat{I}^{(n)}|\hat{I}^{(n)}\rangle} =
 (\tilde{c}_n^{\,*})^2z^{2n-2} + \cdots + \frac{(c_n)^2}{z^{2n+2}}~,
\label{eq:curve4}
\end{align}
up to a change of variables that preserves the SW 1-form. After changing variables as $z \to
z\,(-c_n/\tilde{c}_n^{\,*})^{\frac{1}{2n}}$ and $x\to
x\,(-c_n/\tilde{c}_n^{\,*})^{-\frac{1}{2n}}$, the curve \eqref{eq:curve4}
is re-expressed as $x^2 = (-c_n\tilde{c}_n^{\,*})z^{2n-2} + \cdots
+(-c_n\tilde{c}_n^{\,*})/z^{2n+2}$. Comparing this with \eqref{eq:curve3},
we find $\Lambda_0^2= -c_n\tilde{c}_n^{\,*}$, which
coincides with \eqref{eq:Lambda} up to a numerical factor.

\subsection{Consistency check}
\label{subsec:check}

Since the $(A_1,D_2)$
theory is a theory of free hypermultiplets in the doublet of $U(2)$, one
can perform a consistency check of our proposals \eqref{eq:proposal1}
and \eqref{eq:proposal2} by comparing them with the Nekrasov's formula
for fundamental and anti-fundamental hypermultiplets.

% We here check consistency of our conjecture. In the case of $n=1$, $SU(2)$ gauge theory coupled to two $(A_1,D_{2n})$ theories reduces to $SU(2)$ gauge theory with two fundamental hyper multiplet. The partition function of this theory is respected to be given by
% \begin{align}
%  \mathcal{Z}_{U(2)}^{N_f = 2} = \langle \hat{I}^{(1)}|\hat{I}^{(1)}\rangle~.
% \end{align}
%This equation is rewritten to compare \eqref{eq:Nek1} and
%\eqref{eq:inner2}:
Let us first consider \eqref{eq:proposal1}.
In the case of $n=1$, the irregular state involved in
\eqref{eq:proposal1} satisfies
\begin{align}
 L_2|\hat{I}^{(1)}\rangle &= -c_1^2|\hat{I}^{(1)}\rangle~, 
\\[1mm]
 L_1|\hat{I}^{(1)}\rangle &= 2(Q-c_0)c_1|\hat{I}^{(1)}\rangle~, 
\\
 L_0|\hat{I}^{(1)}\rangle &=
 \left(\Delta_{c_0}+c_1\frac{\partial}{\partial
 c_1}\right)|\hat{I}^{(1)}\rangle~,
\\
 a_1|\hat{I}^{(1)}\rangle &= -ic_1|\hat{I}^{(1)}\rangle~,
\end{align}
together with $a_k|\hat{I}^{(1)}\rangle = 0$ for $k\geq 2$.
% and $a_{k}|\hat{I}^{(1)}\rangle = L_{\ell}|\hat{I}^{(1)}\rangle = 0$
% for $k\geq 2$ and $\ell\geq 3$.
% and
% \begin{align}
%  a_{\ge 2}|\hat{I}^{(1)}\rangle = 0~, \qquad a_1|\hat{I}^{(1)}\rangle = -ic_1|\hat{I}^{(1)}\rangle~.
% \end{align}
% \begin{align}
%  \Lambda^{\frac{Y_1+Y_2}{2}}\mathcal{Z}_{Y_1,Y_2}^{\text{fund}}(a,m) = \langle a;Y_1,Y_2|\hat{I}^{(1)}\rangle,
% \label{eq:fund}
% \end{align}
% where $|\hat{I}^{(1)}\rangle$ which is the $Vir\times H$-irregular state of rank-one satisfies the following equation
These equations are enough to compute the ratio of inner products $\langle
a;Y_1,Y_2|\hat{I}^{(1)}\rangle/\langle a|\hat{I}^{(1)}\rangle$. We then find that\footnote{We checked this equality for
$|Y_1|+|Y_2|\leq 6$.}
\begin{align}
\left(-\frac{c_1}{2}\right)^{-|Y_1|-|Y_2|}\frac{\langle a;Y_1,Y_2|\hat{I}^{(1)} \rangle}{\langle
 a|\hat{I}^{(1)}\rangle} = \mathcal{Z}^\text{fund}_{Y_1,Y_2}(a,m)~,
\label{eq:check1}
\end{align}
where $\mathcal{Z}^\text{fund}_{Y_1,Y_2}$ is the contribution from a
fundamental hypermultiplet of $U(2)$ as reviewed in Appendix
\ref{app:Nek}. % Since the RHS is the contribution to the instanton part
% of the partition function, we normalize the LHS as mentioned at the end
% of Sec.~\ref{subsec:ZAD}.
 The mass parameter $m$ is related to $c_0$ by
\begin{align}
m = c_0-\frac{Q}{2}~,
\label{eq:m1}
\end{align}
which coincides with \eqref{eq:coupling} in the classical limit
$\epsilon_i\to 0$.
We see that \eqref{eq:check1} is perfectly consistent with our proposal
\eqref{eq:proposal1} for $\zeta = -1/2$. 

We also perform the same computation for $\langle
\hat{I}^{(1)}|a;Y_1,Y_2\rangle/\langle \hat{I}^{(1)}|a\rangle$ to find
that\footnote{We also checked this equality for $|Y_1|+|Y_2|\leq 6$.}
\begin{align}
\left(\frac{\tilde{c}_n^{\,*}}{2}\right)^{-|Y_1|-|Y_2|}\frac{ \langle
 \hat{I}^{(1)}|a;Y_1,Y_2\rangle}{\langle \hat{I}^{(1)}|a\rangle} = \mathcal{Z}^\text{anti-fund}_{Y_1,Y_2}(a,\tilde{m})~,
\end{align}
where $\tilde{m}$ is similarly identified as
\begin{align}
 \tilde{m} = \left(\tilde{c}_0-\frac{Q}{2}\right)^*~.
\end{align}
This is also in perfect agreement with our proposal \eqref{eq:proposal2} for
$\zeta = -1/2$.

The above two checks suggest that
the difference between $\mathcal{Z}^{(A_1,D_{2n})}_{Y_1,Y_2}$ and
$\tilde{\mathcal{Z}}^{(A_1,D_{2n})}_{Y_1,Y_2}$ can be regarded as the AD counterpart of
the difference between the fundamental and anti-fundamental hypermultiplets.

% We respect to satisfy \eqref{eq:fund} oder by order of $\Lambda$. We checked to satisfy \eqref{eq:fund} up to 6 instanton. (up to 8 instanton when Liouville charge $Q$ vanishes.)

% *********** WRITE HERE ABOUT THE CONSISTENCY CHECK THAT KIMURA-KUN HAS
%             PERFORMED *************

% \note{Takahiro}{This part must be double checked.}

\section{Application to $(A_3,A_3)$ theory}

\label{sec:A3A3}

% \begin{figure}
%  \begin{center}
% \begin{tikzpicture}[gauge/.style={circle,draw=black,thick,inner sep=0pt,minimum size=8mm},flavor/.style={rectangle,draw=black,thick,inner sep=0pt,minimum size=8mm},AD/.style={rectangle,draw=black,fill=red!20,thick,inner sep=0pt,minimum size=8mm},auto]

% \node[AD] (1) at (-2,0) {\;$(A_1,D_{4})$\;};
% \node[gauge] (2) at (0,0) [shape=circle] {\;$2$\;} edge (1);
% \node[AD] (3) at (2,0)  {\;$(A_1,D_{4})$\;} edge (2);
% \node[flavor] (4) at (0,1.5) {\;1\;} edge (2);

%   \end{tikzpicture}
% \caption{The quiver diagram of $(A_3,A_3)$ AD theory. The left and right
%   boxes stand for two copies of $(A_1,D_4)$ theory while the top box
%   stands for a fundamental hypermultiplet of $SU(2)$. The middle circle
%   stands for $SU(2)$ gauge group coupled to these ``matters.''}
% \label{fig:quiver10}
%  \end{center}
% \end{figure}

In this section, we apply our method to the $(A_3,A_3)$ theory and
compute its partition function. Recall that the $(A_3,A_3)$ theory is
described by the quiver diagram in Fig.~\ref{fig:quiver1}. When the
gauge group is replaced by $U(2)$, the partition function of the theory
is given by
\begin{align}
 \mathcal{Z}_{U(2)} = \mathcal{Z}^{U(2)}_\text{pert}\sum_{Y_1,Y_2}q^{|Y_1|+|Y_2|}
 \mathcal{Z}_{Y_1,Y_2}^\text{vec}(a)\mathcal{Z}_{Y_1,Y_2}^\text{fund}(a,M)\prod_{i=1}^2 \mathcal{Z}_{Y_1,Y_2}^{(A_1,D_4)}(a,m_i,d_i,u_i)~,
\label{eq:Z-A3A3}
\end{align}
where $\mathcal{Z}^{(A_1,D_4)}_{Y_1,Y_2}$ is the contribution of
the $(A_1,D_4)$ theory that we have identified in
Eq.~\eqref{eq:proposal1}, and $\mathcal{Z}^{U(2)}_\text{pert}$ is the
perturbative contribution that makes the series in $q$ start with $1$. The parameters $m_i,d_i$ and $u_i$ are
respectively a
mass parameter, relevant coupling of dimension $1/2$, and the VEV of
Coulomb branch operator of dimension $3/2$ in the $i$-th $(A_1,D_4)$
theory. Since the $SU(2)$ gauge coupling is exactly marginal, the above
expression includes the exponential of the marginal gauge coupling, $q$,
instead of a dynamical scale.

Note that, depending on how the $U(1)$-part of the gauge group
couples to $(A_1,D_4)$, its contribution to the partition function is
$\mathcal{Z}^{(A_1,D_4)}_{Y_1,Y_2}$ or
$\tilde{\mathcal{Z}}^{(A_1,D_4)}_{Y_1,Y_2}$. In this section, we focus on the case
in which both of
the two $(A_1,D_4)$ theories couple to the $U(1)$ in the way
corresponding to $\mathcal{Z}^{(A_1,D_4)}_{Y_1,Y_2}$. Replacing one or
two of $\mathcal{Z}^{(A_1,D_4)}_{Y_1,Y_2}$ with
$\tilde{\mathcal{Z}}^{(A_1,D_4)}_{Y_1,Y_2}$, one would obtain to a different
$\mathcal{Z}_{U(2)}$, which is however expected to give the same $\mathcal{Z}_{SU(2)}$
when the $U(1)$ factor $\mathcal{Z}_{U(1)}$ is removed as in \eqref{eq:U1}.

The factor $\mathcal{Z}_{Y_1,Y_2}^{(A_1,D_4)}$ in Eq.~\eqref{eq:Z-A3A3}
is given by
\begin{align}
 \mathcal{Z}_{Y_1,Y_2}^{(A_1,D_4)}(a,m,d,u) = \left(-\frac{c_2}{2}\right)^{-\frac{|Y_1|+|Y_2|}{2}}\frac{\langle
 a;Y_1,Y_2|\hat{I}^{(2)}\rangle}{\langle a|\hat{I}^{(2)}\rangle}~,
\label{eq:Z-A1D4}
\end{align}
where $m,d$ and $u$ are identified as
\begin{align}
 m = 2c_0 +\frac{c_1^2}{c_2}~,\qquad d =
 \frac{2c_1}{\sqrt{c_2}}~,\qquad u = \frac{2c_0 c_1}{\sqrt{c_2}}-
 \sqrt{c_2}\frac{\partial \mathcal{F}_{(A_1,D_4)}}{\partial c_1}~,
\label{eq:mdu}
\end{align}
with $\mathcal{F}_{(A_1,D_4)}$ being the classical limit of $\log \langle
a|\hat{I}^{(2)}\rangle$. Note that we here set $\zeta = -1/2$ in
\eqref{eq:proposal1} to avoid various numerical factors
in the expressions below. This factor can be generated or absorbed by rescaling $q$ in
the expression \eqref{eq:Z-A3A3}.
The irregular state $|\hat{I}^{(2)}\rangle$ is
characterized by
\begin{align}
 L_4|\hat{I}^{(2)}\rangle &= -c_2^2|\hat{I}^{(2)}\rangle~,
\label{eq:L4}
\\
L_3|\hat{I}^{(2)}\rangle &= -2c_1c_2|\hat{I}^{(2)}\rangle~,
\\
L_2|\hat{I}^{(2)}\rangle &=
 -(c_1^2+c_2(2c_0-3Q))|\hat{I}^{(2)}\rangle~,
\label{eq:L2}
\\
L_1|\hat{I}^{(2)}\rangle &= \left(c_2\frac{\partial}{\partial c_1}
 -2c_1(c_0-Q)\right)|\hat{I}^{(2)}\rangle~,
\\
L_0|\hat{I}^{(2)}\rangle &= \left(\Delta_{c_0} + c_1\frac{\partial}{\partial
 c_1} + 2c_2\frac{\partial}{\partial c_2}\right)|\hat{I}^{(2)}
 \rangle~,
\end{align}
and $a_k|\hat{I}^{(2)}\rangle=-ic_k|\hat{I}^{(2)}\rangle$ for $k=1,2$
together with $a_k|\hat{I}^{(2)}\rangle = 0$ for $k>2$.

To extract the partition function of the $(A_3,A_3)$ theory from
\eqref{eq:Z-A3A3}, one has to remove the contribution of the
$U(1)$-part of the gauge group. As discussed in Sec.~\ref{sec:intro},
this can be done by dividing \eqref{eq:Z-A3A3} by a $U(1)$ factor,
$\mathcal{Z}_{U(1)}$. Therefore the partition function of the
$(A_3,A_3)$ theory is identified as
\begin{align}
 \mathcal{Z}_{(A_3,A_3)} =
 \frac{\mathcal{Z}_{U(2)}}{\mathcal{Z}_{U(1)}}~.
\label{eq:U2/U1}
\end{align}
While it is beyond the scope of this paper to
determine the $U(1)$ factor, we know that $\mathcal{Z}_{U(1)}$ is independent
of the parameter $a$, since $a$ is the VEV of a
scalar in the $SU(2)$ vector multiplet that is neutral under the
$U(1)$. Below, we use this fact and compute the
classical limit of $\mathcal{Z}_{(A_3,A_3)}$.

\subsection{Prepotential}
\label{subsec:prepotential}

Here we consider the classical limit $\epsilon_i\to 0$, and compute the
prepotential of the $(A_3,A_3)$ theory
\begin{align}
 \mathcal{F}^{(A_3,A_3)} \equiv \lim_{\epsilon_i\to 0}
 \left(-\epsilon_1\epsilon_2\log \mathcal{Z}_{(A_3,A_3)}\right)~.
\label{eq:F1}
\end{align}
This prepotential splits into the perturbative and instanton parts as $\mathcal{F}^{(A_3,A_3)} = \mathcal{F}_\text{pert}^{(A_3,A_3)} +
 \mathcal{F}^{(A_3,A_3)}_\text{inst}$, and we are particularly
 interested in the instanton part $\mathcal{F}_\text{inst}^{(A_3,A_3)}$.\footnote{Note that the perturbative part
% our
% classical part \eqref{eq:cl}
 contains the prepotential of the
$(A_1,D_4)$ theories (with their flavor symmetries ungauged).}
% the sum of
% the classical part, 
% \begin{align}
%  \mathcal{F}_\text{cl}^{(A_3,A_3)} = (\log q)a^2 + \sum_{i=1}^2
%  \mathcal{F}_{(A_1,D_4)}(a,m_i,d_i,u_i)~,
% \label{eq:cl}
% \end{align} 
% and the 1-loop correction, $\mathcal{F}^{(A_3,A_3)}_\text{1-loop}$.
%
% in addition to the familiar term $(\log q)a^2$. 
The
instanton part is generally expanded as
\begin{align}
 \mathcal{F}^{(A_3,A_3)}_\text{inst} = \sum_{k=1}^\infty \mathcal{F}_k
 q^k~.
\label{eq:instA3}
\end{align}
Below, we will compute the coefficients, $\mathcal{F}_k$, in this expansion.

To that end, let us first consider
\begin{align}
 \mathcal{F}^{U(2)}\equiv \lim_{\epsilon_i\to 0} \left(-\epsilon_1\epsilon_2\log
 \mathcal{Z}_{U(2)}\right)~,
\label{eq:F2}
\end{align}
which is the prepotential of the theory with the gauge group replaced by
$U(2)$. This prepotential also splits into the perturbative part,
${\displaystyle \lim_{\epsilon_i\to 0} (-\epsilon_1\epsilon_2\log \mathcal{Z}_\text{pert}^{U(2)})}$, and
the instanton part
\begin{align}
\mathcal{F}^{U(2)}_\text{inst} \equiv \lim_{\epsilon_i\to
 0}\left(-\epsilon_1\epsilon_2\log\frac{\mathcal{Z}^{U(2)}}{\mathcal{Z}^{U(2)}_\text{pert}}\right)~.
\label{eq:instU2}
\end{align}
The instanton part \eqref{eq:instU2} is
identical to $\mathcal{F}^{(A_3,A_3)}_\text{inst}$ up to the
contribution of
the $U(1)$ factor. Our
strategy is to compute $\mathcal{F}^{U(2)}_\text{inst}$ using the
formula \eqref{eq:Z-A3A3}, and then
strip off the $U(1)$ factor to obtain $\mathcal{F}_\text{inst}^{(A_3,A_3)}$.

Note 
that the
computation of $\mathcal{Z}_{U(2)}$ via \eqref{eq:Z-A3A3} and \eqref{eq:Z-A1D4} eventually reduces to evaluating 
\begin{align}
 \langle
a|L_{k_p}^{m_p}\cdots L_{k_1}^{m_1}a_{\ell_q}^{n_q} \cdots
a_{\ell_1}^{n_1}|\hat{I}^{(2)}\rangle~,
\label{eq:fragment}
\end{align}
for positive integers $k_i,m_i,\ell_j$ and $n_j$. Using
\eqref{eq:L4}--\eqref{eq:L2} and the fact that $L_k|\hat{I}^{(2)}\rangle
= a_{\ell}|\hat{I}^{(2)}\rangle = 0$ for $k> 4$ and $\ell> 2$, this
computation further reduces to evaluating
\begin{align}
 \langle a|L_1^k|\hat{I}^{(2)}\rangle =
 \left(c_2\frac{\partial}{\partial c_1} - 2c_1(c_0-Q)\right)^k
 \langle a|\hat{I}^{(2)}\rangle~,
\label{eq:fragment2}
\end{align}
where $\langle a| \hat{I}^{(2)}\rangle$ is
the partition function of $(A_1,D_4)$ theory (with its
flavor symmetry ungauged). Therefore, to compute $\mathcal{Z}_{U(2)}$ for general
$\Omega$-background parameters, one needs to know how
$\langle a|\widehat{I}^{(2)}\rangle$ depends on $c_1$.\footnote{The $1/c_1$-expansion of
$\langle a|I^{(2)}\rangle$ was carefully studied in
\cite{Nishinaka:2019nuy}.} However, in the classical limit $\epsilon_i\to
0$, one can skip this procedure. Indeed, recovering the full
$\epsilon_i$-dependence by $c_k\to
c_k/\sqrt{\epsilon_1\epsilon_2}$, we see that in the classical limit
\begin{align}
 \langle a|L_1^k|\hat{I}^{(2)}\rangle = (c_2)^{\frac{k}{2}}(-u)^k
 \langle a |\hat{I}^{(2)}\rangle~,
\label{eq:L1k}
\end{align}
where $u$ is defined by Eq.~\eqref{eq:mdu}. Given
\eqref{eq:L4}--\eqref{eq:L2} and \eqref{eq:L1k}, it is straightforward to
compute the classical limit of $\mathcal{Z}_{U(2)}$, and therefore $\mathcal{F}_\text{inst}^{U(2)}$, order by
order in $q$.
% , where $\mathcal{F}_{(A_1,D_4)}$ in
% \eqref{eq:cl} is identified with ${\displaystyle \lim_{\epsilon_i\to
% 0}(-\epsilon_1\epsilon_2\log \langle a|\hat{I}^{(2)}\rangle )}$.

We now turn to the $U(1)$ factor. While it is generically non-vanishing,
the contribution from the $U(1)$-factor turns out to vanish when all the
dimensionful parameters in four-dimensions, except for $a$ and
$\epsilon_i$, are turned off. Indeed, in the
classical limit, $\mathcal{Z}_{U(1)}\sim
\exp\left(-\frac{1}{\epsilon_1\epsilon_2}\mathcal{F}_{U(1)}\right)$ with
$\mathcal{F}_{U(1)}$ being independent of $\epsilon_i$. When
dimensionful parameters are turned off except for $a$ and $\epsilon_i$ ,
$\mathcal{F}_{U(1)}$ must be proportional to $a^2$ for dimensional reasons. However, as discussed below \eqref{eq:U2/U1},
$\mathcal{Z}_{U(1)}$ must be independent of $a$. This means that the
proportionality constant is zero so that
$\mathcal{F}_{U(1)}=0$. 

Let us now focus on the case in which $M,m_i,d_i$
and $u_i$ in \eqref{eq:Z-A3A3} are all turned off. Then the only
non-vanishing dimensionful parameters are $a$ and $\epsilon_i$. Since the $U(1)$-factor is trivial in
this case, one can identify
\eqref{eq:F1} with \eqref{eq:F2}, and therefore \eqref{eq:instA3} with \eqref{eq:instU2}.
With this identification,
 we finally obtain
\begin{align}
 \mathcal{F}^{(A_3,A_3)}_\text{inst}(q;a) = \left(\frac{1}{4}q^2 + \frac{13}{128}q^4 +
 \frac{23}{384}q^6 + \frac{2701}{65536}q^8 + \cdots \right)a^2~.
\label{eq:F-A3A3}
\end{align}
Remarkably, this expression is closely related to the instanton
part of the prepotential of the $SU(2)$ gauge theory with
four fundamental flavors. Indeed, when all the mass parameters are turned
off, the latter is given by
\begin{align}
 \mathcal{F}_\text{inst}^{N_f=4}(q;a) = \left(\frac{1}{2}q +
 \frac{13}{64}q^2 + \frac{23}{192}q^3 +
 \frac{2701}{32768}q^4 + \cdots \right)a^2~,
\label{eq:F-Nf=4}
\end{align}
as shown in Appendix B.3 of \cite{Alday:2009aq}.
Comparing
\eqref{eq:F-A3A3} and \eqref{eq:F-Nf=4}, we see that
\begin{align}
 2\mathcal{F}_\text{inst}^{(A_3,A_3)}(q;a) =
 \mathcal{F}_\text{inst}^{N_f=4}(q^2,a)~,
\label{eq:surprise}
\end{align}
at least up to $\mathcal{O}(q^8)$.

\subsection{S-duality}
\label{subsec:S-duality}

Here, we show that one can read off the action of  the S-duality group on the $(A_3,A_3)$
theory assuming the remarkable identity \eqref{eq:surprise} extends to
the full prepotential. To that end, let us first give a quick
review of the
S-duality of $SU(2)$ gauge theory with four fundamental flavors. When
the mass parameters are turned off, the full prepotential of this theory
is written as 
\begin{align}
\mathcal{F}^{N_f=4} = (\log q_\text{IR})a^2~,
\label{eq:IR_Nf=4}
\end{align} 
where $q_\text{IR}$ is related to the IR theta angle and electric
coupling by
\begin{align}
 q_\text{IR} = e^{i\theta_\text{IR}- \frac{8\pi^2}{g_\text{IR}^2}}~.
\end{align}
The full prepotential \eqref{eq:IR_Nf=4} is the sum of the instanton
part \eqref{eq:F-Nf=4}
and the perturbative part $(\log q - \log 16)a^2$. This implies that
$q$ and $q_\text{IR}$ are related by \cite{Grimm:2007tm}
\begin{align}
 q =
 \frac{\theta_2(q_\text{IR})^4}{\theta_3(q_\text{IR})^4}~,
\label{eq:UVIR1}
\end{align}
where we used the convention that $\theta_2(q) =
\sum_{n\in\mathbb{Z}}q^{(n-\frac{1}{2})^2}$ and $\theta_3(q) =
\sum_{n\in \mathbb{Z}}q^{n^2}$. The relation \eqref{eq:UVIR1} implies
that  
\begin{align}
 \tau_\text{IR} \equiv \frac{1}{\pi i}\log q_\text{IR} =
 \frac{\theta_\text{IR}}{\pi} + \frac{8\pi i}{g_\text{IR}^2}
\end{align}
is the modulus of the elliptic curve corresponding to the double cover
 of $\mathbb{CP}^1$ with four branch points whose cross-ratio is
 $q$. This elliptic curve is identified as the SW-curve of the theory on
 the Coulomb branch. The curve has a natural $PSL(2,\mathbb{Z})$-action
 generated by
\begin{align}
T: \tau_\text{IR} \to \tau_\text{IR} + 1 ~,\qquad S : \tau_\text{IR} \to
 -\frac{1}{\tau_\text{IR}}~.
\label{eq:TS}
\end{align}
From \eqref{eq:UVIR1}, we see that these $T$ and $S$ transformations are
induced by the following changes of the UV gauge couplings:
\begin{align}
 T:q\to \frac{q}{q-1}~,\qquad S:q\to 1-q~.
\end{align}
Since the SW-curve is invariant under $T$ and $S$, so is the whole BPS
spectrum on the Coulomb branch. It is then natural to expect that the theory is completely
invariant under this $PSL(2,\mathbb{Z})$.\footnote{When mass parameters
are turned on, they are also permuted by the action of
$PSL(2,\mathbb{Z})$.}
This is the famous S-duality of $SU(2)$ gauge theory with four flavors.

% Under $S$, the prepotential $\mathcal{F}^{N_f=4}$ 
% %is invariant under
% % $T$, but 
% transforms as 
% \begin{align}
%  \mathcal{F}^{N_f=4} \to \mathcal{F}^{N_f=4} - a\frac{\partial
%  \mathcal{F}^{N_f=4}}{\partial a}~,
% \end{align} 
% so that $S$ involves the Legendre transform from $a$ to  $a_D \equiv \partial
% \mathcal{F}^{N_f=4}/\partial a$.

Let us now turn back to the $(A_3,A_3)$ theory. Its quiver
 description shown in Fig.~\ref{fig:quiver1} has an obvious similarity
 to the $SU(2)$ gauge theory with four flavors; it has the same gauge
 group with the vanishing $\beta$-function. This similarity has been studied
 carefully in \cite{Buican:2014hfa, DelZotto:2015rca, Cecotti:2015hca}
 to show that the IR physics of
 the $(A_3,A_3)$ theory on its Coulomb branch admits an action of $PSL(2,\mathbb{Z})$ (see \cite{Xie:2016uqq, Xie:2017vaf,
Buican:2017fiq, Xie:2017aqx, Buican:2018ddk} for further studies on this
 new class of $\mathcal{N}=2$ S-dualities). The generalization of the
 $SO(8)$-triality in this case is carefully discussed in \cite{Cecotti:2015hca}. This
 $PSL(2,\mathbb{Z})$-action has only been studied in the IR language
 such as the SW curve, the associated Calabi-Yau three-fold, and the
 spectrum of BPS states on the Coulomb branch. Here we discuss the action of
 $PSL(2,\mathbb{Z})$ on the UV gauge coupling, using the surprising
 relation \eqref{eq:surprise}. 

To that end, let us define the IR gauge coupling $q_\text{IR}$ of the
$(A_3,A_3)$ theory similarly by
\begin{align}
 q_\text{IR} = e^{i\theta_\text{IR} - \frac{8\pi^2}{g_\text{IR}^2}}
\label{eq:qIR-A3A3}
\end{align}
 so that the
full prepotential is written as
\begin{align}
 \mathcal{F}^{(A_3,A_3)} = (\log
q_\text{IR})a^2~.
\label{eq:qIR-def}
\end{align}
Assuming the relation
\eqref{eq:surprise} extends to the full prepotential, one
obtains\footnote{Given the relation \eqref{eq:surprise} for the
instanton part, this assumption is rather mild. Indeed, it only requires
that the perturbative part $\mathcal{F}_\text{pert}  =
\mathcal{F}_\text{cl} + \mathcal{F}_\text{1-loop}$ also satisfies 
\begin{align}
 2\mathcal{F}^{(A_3,A_3)}_\text{pert}(q;a) =
 \mathcal{F}^{N_f=4}_\text{pert}(q^2;a)~.
\label{eq:tobeproven}
\end{align}
Note that, on both sides of the above relation, $\mathcal{F}_\text{pert} = (\log
q + X)a^2$ with $X$ being a constant. Therefore, in proving
\eqref{eq:tobeproven}, all we need to show is the following equality between two constants:
\begin{align}
 2X^{(A_3,A_3)} = X^{N_f=4}~.
\label{eq:1-loop-X}
\end{align}
To prove this, one needs to identify the 1-loop part
$\mathcal{F}_\text{1-loop}$ for $(A_3,A_3)$, which we leave for future
work.
}
\begin{align}
2\mathcal{F}^{(A_3,A_3)}(q;a) = \mathcal{F}^{N_f=4}(q^2;a)~,
\label{eq:surprise2}
\end{align}
which implies that $q_\text{IR}$ is related
to $q$ by
\begin{align}
 q^2 = \frac{\theta_2(q_\text{IR}^2)^4}{\theta_3(q_\text{IR}^2)^4}~.
\label{eq:modular2}
\end{align}
This relation means that 
\begin{align}
 \tau_\text{IR} \equiv \frac{2}{\pi i}\log q_\text{IR} =
 \frac{2\theta_\text{IR}}{\pi} + \frac{16\pi i}{g_\text{IR}^2}
\label{eq:tau2}
\end{align}
is the modulus of the double cover
of $\mathbb{CP}^1$ with four branch points whose cross-ratio is
$q^2$. The
$T$ and $S$ transformations of the S-duality group are identified as
\begin{align}
 T:\tau_\text{IR} \to \tau_\text{IR} + 1~,\qquad S:\tau_\text{IR} \to
 -\frac{1}{\tau_\text{IR}}~.
\label{eq:TS3/2}
\end{align}
% Under $S$, the prepotential transforms as
% \begin{align}
% \mathcal{F}^{(A_3,A_3)} \to \mathcal{F}^{(A_3,A_3)} - a\frac{\partial
%  \mathcal{F}^{(A_3,A_3)}}{\partial a}~,
% \end{align}
% which implies $S$ gives rise to
% the Legendre transform from $a$ to $a_D\equiv \partial
% \mathcal{F}^{(A_3,A_3)}/\partial a$  as expected.

Note that \eqref{eq:modular2} reveals a non-trivial relation between the IR gauge
coupling $q_\text{IR}$ and the UV gauge coupling $q$ of the $(A_3,A_3)$
theory.  Compared to \eqref{eq:UVIR1} for $SU(2)$ gauge theory with four flavors, both
the UV and IR gauge couplings are replaced by their squares here. While the replacement
$q\to q^2$ can be easily understood from the relation
\eqref{eq:surprise2}, the replacement
\begin{align}
 q_\text{IR} \to q_\text{IR}^2
\label{eq:qIR2}
\end{align}
is a bit more non-trivial. We see that this replacement is a consequence
of the factor $2$ in front of $\mathcal{F}^{(A_3,A_3)}$ in
\eqref{eq:surprise2}. Note that \eqref{eq:qIR2} 
is crucial to have the
correct weak-coupling behavior. Indeed, in the weak coupling limit, quantum
corrections to the IR gauge coupling vanish, and therefore we expect
$q=q_\text{IR}$. We see that \eqref{eq:modular2} correctly reduces to
$q\sim q_\text{IR}$ in the weak coupling limit $q_\text{IR} \to 0$
if \eqref{eq:qIR2} is simultaneously performed with $q\to q^2$.

From the above discussion, we see how $S$ and $T$ act on
the UV gauge coupling of the $(A_3,A_3)$ theory. Indeed, combining \eqref{eq:modular2} and
\eqref{eq:tau2}, we see that \eqref{eq:TS3/2} corresponds
to
\begin{align}
 T:\; q^2 \to \frac{q^2}{q^2-1}~,\qquad S:\; q^2\to 1-q^2~.
\label{eq:TS2}
\end{align}
One can explicitly check that 
\eqref{eq:F-A3A3} combined with the classical part
$\mathcal{F}^{(A_3,A_3)}_\text{cl} = (\log q)a^2$ is invariant under $T$.

\subsection{Peculiarity of $T$}

\label{subsec:peculiarT}

% From \eqref{eq:qIR-def}, we find that
% the IR theta angle $\theta_\text{IR}$ and Yang-Mills coupling
% $g_\text{YM}$ are related to $q_\text{IR}$ by
% \begin{align}
%  q_\text{IR} = e^{i \theta_\text{IR} - \frac{8\pi^2}{g_\text{IR}^2}}~.
% \end{align}
From \eqref{eq:tau2}, we see that our $T$-transformation,
$\tau_\text{IR} \to \tau_\text{IR}+1$, corresponds to
\begin{align}
 \theta_\text{IR} \to \theta_\text{IR} + \frac{\pi}{2}~.
\label{eq:T_for_theta}
\end{align}
This is remarkable since this means that $T$ maps the 
monopole of the minimal magnetic charge to a dyon that has {\it half} the
electric charge of fundamental quark. This is not possible if this
``electric charge'' is a charge associated with the unbroken $U(1)\subset SU(2)$ gauge
group on the Coulomb branch, since in that case the minimal electric
charge is the charge of fundamental quark. Therefore,
the ``electric charge'' here is not simply associated with the unbroken $U(1)\subset SU(2)$.
As we will argue in Sec.~\ref{sec:Conclusions}, the ``electric charge'' here
is interpreted as a linear combination of the electric charge associated
with $U(1)\subset SU(2)$ and those arising from the $(A_1,D_4)$ theories.

% We will argue in
% Sec.~\ref{sec:Conclusions} that this is
% indeed consistent with the result of \cite{Cecotti:2015hca}.

In the rest of this sub-section, we show that the $T$ transformation \eqref{eq:T_for_theta} is also consistent with the SW curve of $(A_3,A_3)$ theory. To that end, first recall that the SW curve of this theory is written as \cite{Buican:2014hfa} 
\begin{align}
 0 &= x^4 +\mathtt{q}x^2z^2 + z^4 + c_{3,0}x^3 + c_{0,3}z^3 + c_{2,0}x^2
 + mxz + c_{0,2}z^2 + c_{1,0}x + c_{0,1}z + c_{0,0}~,
\label{eq:curve}
\end{align}
where $\mathtt{q}$ is a function of $q_\text{IR}$, $c_{3,0}$ and
$c_{0,3}$ are relevant couplings of dimension $1/2$,
$c_{2,0},\,c_{0,2}$ and $m$ are mass parameters,
and $c_{1,0},\,c_{0,1}$ and $c_{0,0}$ parameterize the Coulomb branch moduli
space. The SW 1-form is given by $xdz$. It was
shown in \cite{Buican:2014hfa} that the above curve and 1-form are invariant
under two transformations $\tilde{S}$
 and $\tilde{T}$, which were interpreted as two independent S-dual
 transformations. In particular, $\tilde{S}$ acts on the marginal gauge
 coupling as\footnote{The $\tilde{T}$-transformation acts on the gauge
 coupling as $\tilde{T}: \mathsf{q}\to
 \frac{12-2\mathsf{q}}{2+\mathsf{q}}$. Our $T$ and $S$ correspond to
 $\tilde{S}$ and $\tilde{T}$ in \cite{Buican:2014hfa}, respectively.}
\begin{align}
 \tilde{S}:\; \mathsf{q} \to -\mathsf{q}~.
\label{eq:tildeS}
\end{align}

Below, we show that \eqref{eq:tildeS} is identical to our
$T$-transformation \eqref{eq:T_for_theta}, near a cusp on the
conformal manifold.\footnote{Here, the ``conformal manifold'' is defined
as the space of values of exactly marginal couplings in the theory.}
As shown
in \cite{Buican:2014hfa}, $\mathsf{q}\to \infty$ corresponds to a
weak-coupling cusp on the conformal manifold, where $M\equiv m/\mathsf{q}$
and $u\equiv c_{0,0}/\mathsf{q}$ are respectively
identified as the mass of the fundamental hypermultiplet and the VEV of
the Coulomb branch operator in the vector multiplet, in the quiver
description in Fig.~\ref{fig:quiver1}.\footnote{There are also
other cusps at $\mathsf{q} \to \pm 2$, where the parameters in the SW
curve have different interpretations. In particular, the mass of the
fundamental hypermultiplet is identified with some linear combination of
$m, c_{2,0}$ and $c_{0,2}$.} We focus on this cusp,
 and turn on an infinitely large mass, $M$, for the fundamental hypermultiplet so
that it decouples from the theory in the infrared. When the
hypermultiplet decouples, the theory reduces to a non-conformally gauged
AD theory described in Fig.~\ref{fig:quiver2}. To realize this limit at
the level of the SW curve, one needs to take $\mathsf{q}\to \infty$
simultaneously with $M\to \infty$. Indeed, if we take $M\to \infty$ with
$\mathsf{q}$ kept fixed, some periods of the
curve are divergent. It turns out that all the periods of the curve
remain finite when one takes $M\to \infty$ and $\mathsf{q}\to \infty$
with 
\begin{align}
 \Lambda \equiv \frac{M}{\sqrt{\mathsf{q}}}
\label{eq:Lambda-finite}
\end{align}
kept fixed.
This $\Lambda$ is then identified as
a dynamical scale of the resulting theory. The curve correctly reduces to the curve of the IR theory \eqref{eq:curve3} (for $n=2$) in the limit $M,\mathsf{q}\to \infty$ 
with \eqref{eq:Lambda-finite} kept finite.
\footnote{One can show this explicitly as follows. As show in
\cite{Buican:2014hfa}, near the cusp $\mathsf{q}\to\infty$, $C_{3,0} \equiv
\mathsf{q}^{-\frac{1}{4}}c_{3,0}$ and $C_{0,3}\equiv \mathsf{q}^{
-\frac{1}{4}} c_{0,3}$ are identified with the (correctly-normalized) relevant couplings of
dimension $1/2$, $C_{2,0}\equiv \mathsf{q}^{-\frac{1}{2}}c_{2,0}$ and
$C_{0,2}\equiv \mathsf{q}^{-\frac{1}{2}}c_{0,2}$ are mass deformation
parameters, and $C_{1,0}\equiv \mathsf{q}^{-\frac{3}{2}}c_{1,0}$ and
$C_{0,1}\equiv \mathsf{q}^{-\frac{3}{4}}c_{0,1}$ are the VEVs of Coulomb
branch operators, in the $(A_1,D_4)$ sectors. In terms of these
variables, the curve of the $(A_3,A_3)$ theory is written as 
$0 = x^4 + \mathsf{q}x^2z^2 + z^4 + \mathsf{q}^{\frac{1}{4}}C_{3,0}x^3
 + \mathsf{q}^{\frac{1}{4}}C_{0,3}z^3 +
 \mathsf{q}^{\frac{1}{2}}C_{2,0}x^2  + \mathsf{q}Mxz+
 \mathsf{q}^{\frac{1}{2}}C_{0,2}z^2 + \mathsf{q}^{\frac{3}{4}}C_{1,0}x +
 \mathsf{q}^{\frac{3}{4}}C_{0,1}z + \mathsf{q}u$. We here define $X\equiv
i(\sqrt{z}x^{\frac{3}{2}} +
 \frac{1}{2}\sqrt{q}\Lambda\sqrt{x/z}),\,Z\equiv \sqrt{z/x}$ and $U
 \equiv u - \frac{q\Lambda^2}{4}$, and take the limit $M,\mathsf{q}\to
 \infty$ with $\Lambda,U,C_{i,j}$ and $(X,Y)$ kept finite. Then the
 curve reduces to
% \begin{align}
%  0 &= (XZ-\frac{\sqrt{q}\Lambda}{2})^2\left(\frac{1}{qZ^4} + 1 + \frac{Z^4}{q}\right) +
%  (XZ-\frac{\sqrt{q}\Lambda}{2})^{\frac{3}{2}}\left(\frac{C_{3,0}}{q^{\frac{3}{4}}}\frac{1}{Z^3} + \frac{C_{0,3}}{q^{\frac{3}{4}}}Z^3\right)
%  \nonumber\\
% &\qquad + (XZ-\frac{\sqrt{q}\Lambda}{2})\left(\frac{C_{2,0}}{q^{\frac{1}{2}}Z^2} + \sqrt{q}\Lambda  +
%  \frac{C_{0,2}}{q^{\frac{1}{2} 
% }}Z^2\right)+ (XZ-\frac{\sqrt{q}\Lambda}{2})^{\frac{1}{2}}\left(\frac{C_{1,0}}{q^{\frac{1}{4}}}\frac{1}{Z} + \frac{
% C_{0,1}}{q^{\frac{1}{4}}}Z\right) + u
% \end{align}
\begin{align}
X^2 &=  \tilde{\Lambda}^2Z^2 + \tilde{\Lambda}^{\frac{3}{2}}C_{0,3}Z +
 \tilde{\Lambda}C_{0,2}+ \frac{\tilde{\Lambda}^{\frac{1}{2}}C_{0,1}}{Z}
 + \frac{U}{Z^2} + \frac{\tilde{\Lambda}^{\frac{1}{2}}C_{1,0}}{Z^3} +
 \frac{\tilde{\Lambda}C_{2,0}}{Z^4} +
 \frac{\tilde{\Lambda}^{\frac{3}{2}}C_{3,0}}{Z^5}+
 \frac{\tilde{\Lambda}^2}{Z^6}~,
\end{align}
where $\tilde{\Lambda}\equiv -\Lambda/2$. The SW 1-form is written as
$-\frac{3}{2}iXdZ$ up to exact terms. This curve is
precisely identical to \eqref{eq:curve3} for $n=2$. The 1-form is also
identical up to a prefactor that can be absorbed by rescaling
dimensionful parameters and $X$. This implies that
\eqref{eq:Lambda-finite} is the correct identification of the IR
dynamical scale.
}

Note here that, by standard arguments, the dynamical
scale $\Lambda$ of the mass-deformed theory and the gauge coupling $e^{i\theta_\text{IR} -
\frac{8\pi^2}{g_\text{IR}^2}}$ of the conformal theory are related by 
\begin{align}
 \left(\frac{\Lambda}{M}\right)^{b_0} = e^{i\theta_\text{IR}
 -\frac{8\pi^2}{g_\text{IR}^2}}~,
\label{eq:Lambda-finite2}
\end{align}
where $b_0$ is the coefficient of the one-loop $\beta$-function of the
IR theory. Since $b_0 = 1$ in our case, \eqref{eq:Lambda-finite} and
\eqref{eq:Lambda-finite2} imply that
\begin{align}
 \mathsf{q}  = e^{-2i\theta_\text{IR} + \frac{16\pi^2}{g_\text{IR}^2}}~.
\end{align}
From this relation, it is now clear that the $\tilde{S}$ transformation \eqref{eq:tildeS}
is precisely identical to our $T$-transformation \eqref{eq:T_for_theta}. Therefore,
our $T$-transformation \eqref{eq:T_for_theta} is perfectly consistent
with the earlier analysis of the S-duality using the SW-curve.

\subsection{Turning on couplings, masses and VEVs}
\label{subsec:massive-def}

In the previous sections, we have identified the $PSL(2,\mathbb{Z})$-action in the case
of vanishing dimensionful parameters except for $\epsilon_i$ and
$a$. This action has been interpreted as corresponding to the symmetry of the
SW-curve studied in \cite{Buican:2014hfa}. Since this symmetry of the
curve extends to the case of generic values of dimensionful parameters, we
expect that the $PSL(2,\mathbb{Z})$-action on the partition function has
a similar extension for non-vanishing relevant couplings, masses and
VEVs of Coulomb branch operators. 
 In this sub-section, we discuss such an extension of the $PSL(2,\mathbb{Z})$-action.%  to the case in which the relevant
% couplings, masses and VEVs of Coulomb branch operators are turned
% on. Since the symmetry of the SW-curve admits such an extension, 

The instanton part of the prepotential for generic values of
dimensionful parameters is expanded as
\begin{align}
\mathcal{F}^{(A_3,A_3)}_\text{inst} &= \sum_{k=-1}^\infty
 \mathcal{F}_{2k}(q,M,\{m_i\},\{u_i\},\{d_i\})a^{-2k}~.
% \nonumber\\
% &=
%  \mathcal{F}_{-2}(q)\,a^2 + \mathcal{F}_{0}(q,m_i,M,u_i,d_i) +
%  \mathcal{F}_2(q,m_i,M,u_i,d_i)\frac{1}{a^2} +
%  \mathcal{O}\left(\frac{1}{a^4}\right)~.
\label{eq:massiveF}
\end{align}
% Note that
% $\mathcal{F}_{k>0}$ is not affected by the $U(1)$-factor since the
% $U(1)$-factor is independent of $a$. 
Recall that $T$
and $S$ act on the UV gauge coupling as in \eqref{eq:TS2}.
Since $S$ is non-perturbative in $q$, it is hard to understand how
 \eqref{eq:massiveF} behaves under $S$ using our order-by-order computation.
We therefore focus on the $T$-transformation below.
%  A possible
% direction for a check of $S$ will be discussed in Sec.~\ref{sec:Conclusions}.

In the case of $d_i=m_i=u_i=M=0$, \eqref{eq:massiveF} reduces to
\begin{align}
\mathcal{F}_{-2}(q)a^2~,
\label{eq:genus0}
\end{align}
whose explicit expression is shown in Eq.~\eqref{eq:F-A3A3}. 
When combined with the perturbative part, this is
invariant under $T:q^2 \to q^2/(q^2-1)$. 

When $d_i,m_i,u_i$ and $M$ are
turned on, \eqref{eq:genus0} receives corrections from $\mathcal{F}_{2k}$ for all $k\geq 0$. We evaluated
these corrections using our formula \eqref{eq:Z-A1D4} to find
that $\mathcal{F}_{-2}$ and $\mathcal{F}_{k>0}$ are all
invariant under
\begin{align}
 q&\to \frac{iq}{\sqrt{1-q^2}}~,\quad d_1 \to \frac{d_1+q
 d_2}{\sqrt{1-q^2}}~,\quad d_2 \to  id_2~,\quad m_2  \to
 -m_2~,\quad u_2 \to -iu_2~,
\label{eq:T-massive}
\end{align}
with the other parameters kept fixed.\footnote{Note that
$\mathcal{F}_0$ cannot be evaluated without identifying the
$U(1)$-factor, and therefore we leave
the computation of $\mathcal{F}_0$ for future work. The other terms,
$\mathcal{F}_{-2}$ and $\mathcal{F}_{k>0}$, are not affected by the $U(1)$-factor.}
% Here, $d_2,m_2$ and $u_2$ are parameters of the second $(A_1,D_4)$ sector, while
% $d_1$ is the relevant coupling of the first $(A_1,D_4)$ sector; the mass
% $m_1$ and the Coulomb branch parameter $u_1$ of the first $(A_1,D_4)$
% sector are kept fixed.
 We checked this invariance
up to a very high order of $q$. First few terms in the $q$-series
of $\mathcal{F}_{-2},\mathcal{F}_{2}$ and $\mathcal{F}_4$ are shown in appendix
\ref{app:F2}. Given this invariance, we
interpret \eqref{eq:T-massive} as an extension of $q^2\to q^2/(q^2-1)$
to the case of non-vanishing $d_i,m_i,u_i$ and $M$.

% ******* and expect that
% \eqref{eq:T-massive} is a generalization of $T$  to
% the case with non-vanishing $d_i,m_i,u_i$ and $M$.*********
 
It turns out that \eqref{eq:T-massive} is consistent with the symmetry
of the SW curve, at least in the weak coupling limit. To see this,
recall that our $T$-transformation 
is identified with $\tilde{S}$ discussed in \cite{Buican:2014hfa}. This $\tilde{S}$ induces $\mathsf{q}\to -\mathsf{q}$, as
reviewed already, and also changes the other parameters in the SW curve as
\begin{align}
 &c_{3,0} \to -e^{\frac{3\pi i}{4}}c_{3,0}~,\quad c_{0,3}\to
 -e^{-\frac{3\pi i}{4}}c_{0,3}~,\quad c_{2,0} \to -ic_{2,0}~,\quad
 c_{0,2}\to ic_{0,2}~,
\nonumber\\
&m \to -m~,\quad c_{1,0}\to -e^{\frac{\pi
 i}{4}}c_{1,0}~,\quad c_{0,1}\to -e^{-\frac{\pi i}{4}}c_{0,1}~,\quad
 c_{0,0}\to -c_{0,0}~.
\label{eq:tilde-S}
\end{align}
It is straightforward to show that the curve \eqref{eq:curve} and the
SW 1-form is
invariant under these transformations.
Since
$q^2 \to q^2/(q^2-1)$ is already identified with $\mathsf{q} \to -\mathsf{q}$, 
\eqref{eq:tilde-S} is expected to be equivalent to
\eqref{eq:T-massive}. One can show this equivalence explicitly at the
weak-coupling cusp $\mathsf{q}=\infty$.
As shown in
\cite{Buican:2014hfa}, in the limit $\mathsf{q}\to\infty$, the coefficients $c_{i,j}$ must
be appropriately renormalized so that physical
quantities remain finite. In terms of our $d_i,m_i,u_i,M$ and $u$, such
a renormalization is expressed as
\begin{align}
 &c_{3,0} = \mathtt{q}^{\frac{1}{4}}d_1~,\quad c_{0,3} =
 \mathtt{q}^{\frac{1}{4}}d_2~,\quad c_{2,0} =
 \mathtt{q}^{\frac{1}{2}}m_1~,\quad c_{0,2} =
 \mathtt{q}^{\frac{1}{2}}m_2~,\quad m = \mathtt{q}M~,
\nonumber \\
&c_{1,0} = \mathtt{q}^{\frac{3}{4}}u_1~,\quad c_{0,1}= \mathtt{q}^{\frac{3}{4}}u_2~,\quad
 c_{0,0} = \mathtt{q} u~.
\label{eq:relation}
\end{align}
From \eqref{eq:relation} and \eqref{eq:tilde-S}, we see that $\tilde{S}$
is equivalent to $\mathsf{q}\to \exp(-\pi i)\mathsf{q}$ combined with
\begin{align}
d_2 \to
 id_2~,\qquad m_2 \to -m_2~,\qquad u_2\to -iu_2~,
\end{align}
where $d_1,m_1,u_1,M$ and $u$ are kept fixed. Since $\mathsf{q}\to
\exp(-\pi i)\mathsf{q}$ is identified with $q^2\to q^2/(q^2-1)$, this is identical
to \eqref{eq:T-massive} at the leading
order of $q$.
The reason that we only see the leading order terms
is that we are taking the weak-coupling limit $\mathtt{q}\to \infty$ here,
corresponding to $q\to 0$. 
Thus, we have shown that our \eqref{eq:T-massive}
is equivalent to \eqref{eq:tilde-S} in the weak-coupling
limit.

Note that the invariance of $\mathcal{F}_\text{inst}^{(A_3,A_3)}$ under \eqref{eq:T-massive} strongly constrains
the possible form of $\mathcal{F}_{k> 0}$. In particular, when combined
with the trivial symmetry under exchanging two $(A_1,D_4)$ theories in
the quiver diagram,
imposing this invariance determines $\mathcal{F}_{k>0}$ up to
some $T$-invariant functions for every $k$. We explicitly show this for
$\mathcal{F}_{2}$ in appendix \ref{app:F2}.

Note also that our check of the invariance under \eqref{eq:T-massive}
is only for the instanton part of
the prepotential. For the full prepotential to be invariant under \eqref{eq:T-massive},
the perturbative part must also be invariant by itself. One can
show that the perturbative part is invariant under \eqref{eq:T-massive}
when $q=0$, but its extension to the case of $q\neq0$ is not
straightforward. If \eqref{eq:T-massive} also preserves the perturbative
part for general $q$, it is natural to identify \eqref{eq:T-massive} as the
$T$-transformation for generic values of $d_i,m_i$ and $u_i$. If
the invariance under \eqref{eq:T-massive} does not extend to the perturbative part, \eqref{eq:T-massive}
is a new symmetry that only preserves the instanton part. We leave a
detailed study of these two options for future work.

%The fact that $\mathcal{F}_{k>0}$ is invariant under
%\eqref{eq:T-massive} is then a
%highly non-trivial consistency check for our proposal \eqref{eq:Z-A3A3} and
%\eqref{eq:Z-A1D4}.

\section{Conclusions and discussions}
\label{sec:Conclusions} 

In this paper, we have proposed a Nekrasov-type formula for the
instanton partition function of four-dimensional $\mathcal{N}=2\;\,
U(2)$ gauge theories coupled to $(A_1,D_{2n})$ Argyres-Douglas
theories, by extending the generalized AGT correspondence to the case of
$U(2)$ gauge group. We have defined irregular states of the
direct sum of Virasoro and Heisenberg algebras, and then identified 
the contribution of the $(A_1,D_{2n})$ theory at each fixed points on
the $U(2)$
instanton
moduli space, as in
\eqref{eq:proposal1} and \eqref{eq:proposal2}. Here, 4d and 2d
parameters are related by \eqref{eq:coupling}.

As an application, we evaluate the instanton partition function of the
$(A_3,A_3)$ theory. While this partition function cannot be directly
evaluated by the AGT correspondence, we have computed it using our
formula above.
Our result shows that, when some
parameters are turned off, the instanton part of the prepotential of
$(A_3,A_3)$ is in a surprising relation to that of $SU(2)$ gauge theory with four flavors,
as shown in \eqref{eq:surprise}. In particular, the two prepotentials
are related by the following replacement of the UV gauge coupling:
\begin{align}
 q\to q^2~.
\label{eq:qq2}
\end{align}
From this relation, we have read off the
action of the S-duality group on the UV gauge coupling $q$. We have also
discussed its possible extension to dimensionful parameters.

Here, we give a brief comment on our peculiar $T$-transformation. As
shown in Sec.~\ref{subsec:peculiarT}, the $T$-transformation,
$\tau_\text{IR} \to \tau_\text{IR} + 1$, corresponds to 
\begin{align}
 \theta_\text{IR} \to \theta_\text{IR} + \frac{\pi}{2}~,
\label{eq:T-again}
\end{align}
and therefore maps the monopole of the minimal magnetic
charge to a dyon that has half the electric charge of
fundamental quark. This
implies that the ``electric charge'' here is not simply the electric charge
associated with the $SU(2)$ vector multiplet in
Fig.~\ref{fig:quiver1}. Instead, this electric charge is
interpreted as a linear combination of the electric charge associated with the $SU(2)$
vector multiplet sector and those associated with the $(A_1,D_4)$ sectors. Indeed, it was shown in \cite{Cecotti:2015hca}
that $PSL(2,\mathbb{Z})$ naturally acts on a modified electro-magnetic
charge lattice where the ``electric charge'' is such a linear
combination of charges arising from the three sectors.\footnote{The ``magnetic charge'' here
is simply associated with the $SU(2)$ vector multiplet without mixing
with those arising from the $(A_1,D_4)$ sectors.} This linear combination naturally arises
when constructing the $(A_3,A_3)$ theory as the IR limit of type IIB
string theory on a Calabi-Yau singularity.
With this modified charge lattice, the minimal
electric charge is now half the charge of
fundamental quark,\footnote{This can be seen from the discussions 
%in
%Sec.~2.2 and 2.3 of 
around Eq.~(2.64) of \cite{Cecotti:2015hca},
where $\mathtt{deg}\, X$ and $\mathtt{rank}\, X$ are respectively the
(modified) electric and magnetic
charges naturally acted on by $PSL(2,\mathbb{Z})$. Their discussions
here apply to a larger class of theories labeled by $p=2,3,4$ and $6$, and the $(A_3,A_3)$ theory
corresponds to the $p=4$ case. As the authors explain there,
$\mathtt{deg}\, X$ is quantized in the unit of $1/p = 1/4$ when it is
normalized so that the W-boson has charge $1$. This means that the
minimal value of the modified electric charge, $\mathtt{deg}\, X$, is half the charge of
fundamental quark.}%  which is consistent with our result in
% Sec.~\ref{subsec:peculiarT}. 
which is consistent with our T-transformation \eqref{eq:T-again}.

There are obviously many interesting open problems. We list some of them below.
\begin{itemize}
\item While we have focused on the $(A_1,D_{k})$ theories for
      even $k$, there are also those theories for odd
      $k$. Since they also have $SU(2)$ flavor symmetry that can be
      gauged, one can consider the generalization of our
      work to this latter class of theories. One difficulty is that
      the relevant irregular state in this case
      cannot be obtained in a colliding limit. Therefore, it is not
      straightforward to derive the action of the Heisenberg algebra on
      the irregular state. 

 \item It is also interesting to generalize our work to $SU(N)$ gauge theories coupled to
       AD theories. For that, we need a $U(N)$-version of the
       generalized AGT correspondence. A careful study on the $SU(3)$-version of the generalized AGT
       correspondence has been carried out in \cite{Kanno:2013vi}.

 \item It would be interesting to study the Nekrasov-Shatashvili
       limit \cite{Nekrasov:2009rc} of the $\Omega$-deformed $(A_3,A_3)$
       theory. In this limit, combining the
       results of \cite{Ito:2017iba,Ito:2018hwp,Ito:2019twh,Ito:2020lyu}
       with our formula, one can evaluate the deformed prepotential
       of the
       $(A_3,A_3)$ theory including both the perturbative and instanton parts.

 \item The uplift of the AGT correspondence to five dimensions are known
       \cite{Awata:2009ur, Awata:2010yy, Yanagida:2010vz, Taki:2014fva, Mitev:2014isa,
       Isachenkov:2014eya, Ohkubo:2015roe, Bourgine:2016vsq,
       Pasquetti:2016dyl, Negut:2016dxr}. It would be interesting to
       search for a 5d
       uplift of our results. That would shed some light on the relation
       between the 4d $D_{2n}(SU(2))$ theory and the 5d $\hat{D}_{2n}(SU(2))$
       theory \cite{DelZotto:2015rca, Hayashi:2017jze}. More
       generally, it would
       be interesting to study how our results are phrased in terms
       of the $W_{1+\infty}$ algebra and the DIM algebra
       along the lines of \cite{Kanno:2011qv, Awata:2011dc, Kanno:2013aha,
       Bourgine:2015szm, Mironov:2016yue,
       Awata:2016riz,Awata:2016mxc,Awata:2016bdm, Awata:2017cnz,
        Bourgine:2017jsi,  Bourgine:2017rik,Awata:2017lqa,Sasa:2019rbk}.

 \item It was shown in \cite{Buican:2019kba} that the Schur indices of the
       $(A_3,A_3)$ theory and $SU(2)$ gauge theory with four flavors are
       related by a change of variables involving
\begin{align}
 q\to q^2~,
\end{align}
where $q$ is now the superconformal fugacity of the index. Although this
       $q$ is different from our $q$ in \eqref{eq:qq2}, it would be
       very interesting to study how these two peculiar relations are
       connected. 

 \item Our discussion in the previous bullet suggests that the
       Schur index and the prepotential might generally be related. Since the
       Schur index is identified as the
       character of the associated chiral algebra \cite{Beem:2013sza},
       this then suggests a possible connection between the chiral
       algebra and the prepotential. Such a connection has already been
       suggested
       in the application of the ODE/IM correspondence to quantum SW-curves \cite{Ito:2017ypt}.
       It would be interesting to study
       this connection further.

\item  A replacement similar to \eqref{eq:qq2} arises when considering
       the Nekrasov partition functions of $SO/Sp$ gauge
       theories in connection to $SU$ gauge
       theories \cite{Hollands:2010xa, Hollands:2011zc}. This seems to suggest that a $\mathbb{Z}_2$-orbifolding
       plays some role similarly in our case. It would be interesting to study
       this possibility further.

\end{itemize}

\section*{Acknowledgements}

We would like to thank Matthew Buican, Katsushi Ito, Kazunobu Maruyoshi, Hiraku
Nakajima, Satoshi
Nawata, Takuya Okuda, Jaewon Song, Yuji Tachikawa, and Wenbin
Yan for various illuminating discussions. We would like to particularly
thank Kazunobu Maruyoshi for drawing our attention to the reference
\cite{Alba:2010qc} when one of us (T. N.) had a discussion with him during the international
conference ``KEK Theory Workshop 2018.'' T.~N. also thanks KEK
theory group for hosting the wonderful workshop with various
opportunities of discussions. T.~N.'s research is partially supported by
JSPS Grant-in-Aid for Early-Career Scientists 18K13547. The work of T.~U. is supported by Grant-in-Aid for JSPS Research Fellows 19J11212.

\appendix

\section{Nekrasov's formula}
\label{app:Nek}

Here, we list formulae for contributions
to the Nekrasov
partition function of $U(2)$ gauge theories. 
As discussed around
Eq.~\eqref{eq:Nek1}, the instanton part of the partition function is
written as a sum over pairs of Young diagrams, $(Y_1,Y_2)$. For each
$(Y_1,Y_2)$, the $U(2)$ vector multiplet contributes the following factor
\begin{align}
 \mathcal{Z}^\text{vec}_{Y_1,Y_2}(a) &\equiv \prod_{i,j=1}^2 \prod_{s\in
 Y_i} \frac{1}{-E_{Y_i,Y_j}(a_i-a_j,s)+\epsilon_1+\epsilon_2}\prod_{t\in
 Y_j} \frac{1}{E_{Y_j,Y_i}(a_j-a_i,t)}~,
\end{align}
where $a_1=-a_2 = a$ corresponds to the Coulomb branch parameter, and
\begin{align}
  E_{Y_1,Y_2}(a,s) &\equiv a-\epsilon_1 L_{Y_2}(s) +
 \epsilon_2(A_{Y_1}(s)+1)~.
\label{eq:E}
\end{align}
Here $L_Y(s)$ and $A_Y(s)$ are defined as follows. 
For a Young diagram $Y = \{\lambda_1 \geq \lambda_2\geq
\cdots \}$, we denote its transpose by $\{\lambda_1'\geq \lambda_2'\geq
\cdots\}$. We also denote  by $s=(i,j)$  a position in a Young diagram
$Y$. Then the leg-length and arm-length are defined by $L_Y(s)\equiv
\lambda_j'-i$ and $A_Y(s)\equiv \lambda_i-j$, respectively.

The contribution from a fundamental hypermultiplet of $U(2)$ is given by
\begin{align}
\mathcal{Z}^\text{fund}_{Y_1,Y_2}(a,m) &\equiv \prod_{i=1}^2\prod_{s\in
 Y_i}\left(\phi(a_i,s)-m+\epsilon_1+\epsilon_2\right)~,
\end{align}
where 
\begin{align}
\phi(a,s) &\equiv a + \epsilon_1(i-1) + \epsilon_2(j-1)~,
\end{align}
and $m$ is the mass of the hypermultiplet.
The contribution from a bi-fundamental hypermultiplet of $U(2)\times
U(2)$ is written as
\begin{align}
 \mathcal{Z}^\text{bifund}_{Y_1,Y_2;W_1,W_2}(a,b,\alpha) &\equiv
 \prod_{i,j=1}^2\prod_{s\in
 Y_i}\left(E(a_i-b_j,Y_i,W_j,s)-\alpha\right)
\nonumber\\
&\qquad \times \prod_{t\in
 W_j}\left(\epsilon_1+\epsilon_2-E(b_j-a_i,W_j,Y_i,t)-\alpha\right)~,
\end{align}
where $(Y_1,Y_2)$ and $(W_1,W_2)$ correspond to torus fixed points on
the instanton moduli spaces of two $U(2)$ gauge groups, and $a$ and $b$
are the Coulomb branch parameters of these two $U(2)$. The extra
parameter, $\alpha$, stands for the mass parameter of the bi-fundamental hypermultiplet.

\section{Orthogonal Basis}
\label{app:basis}

Here we list the first few examples of states $|a;Y_1,Y_2\rangle$ in the basis of the highest weight
module of $Vir\oplus H$ that
we have used in Sec.~\ref{sec:Nek-AD}. These states are defined as
solutions to \eqref{eq:basis}, and were first found in
\cite{Alba:2010qc}. For ease of reference, we here denote the highest weight
by $P$ instead of $a$, as in \cite{Alba:2010qc}. 

To describe these states, we first decompose the space of states by the level
of descendants. Here, the ``level'' is the sum of the levels of the
Virasoro part and the Heisenberg part, so that $L_{-k_1}^{m_1}\cdots
L_{-k_p}^{m_p}a_{-\ell_1}^{n_1}\cdots a_{-\ell_q}^{n_q}|P\rangle$ has
level $\sum_{i=1}^p m_ik_i + \sum_{j=1}^q n_j\ell_j$. The basis $\{|P;Y_1,Y_2\rangle\}$ found in
\cite{Alba:2010qc} is such that $|P;Y_1,Y_2\rangle$ for $|Y_1|+|Y_2| =
k$ span the space of level-$k$ descendants. Recall here that we are rescaling dimensionful
parameters so that $\epsilon_1\epsilon_2=1$. Therefore we set
$\epsilon_1 = 1/\epsilon_2 = \mathtt{b}$ when computing the RHS of
\eqref{eq:basis}. In this case, the Liouville charge is written as $Q =
\mathtt{b} + 1/\mathtt{b}$. Below, we list the first few
examples of $|P;Y_1,Y_2\rangle$:
\begin{align}
|P;\emptyset,\emptyset\rangle &= |P\rangle~,
\\[1mm]
|P; {\tiny\yng(1)},\emptyset\rangle &=\left(-i(\mathtt{b}+\mathtt{b}^{-1}+2P)a_{-1} - L_{-1}\right)|P\rangle~,
\\[2mm]
|P;{\tiny\yng(1,1)},\emptyset\rangle &= 
\Big( -i\mathtt{b}^{-1}(\mathtt{b}+\mathtt{b}^{-1}+2P)(\mathtt{b}+2\mathtt{b}^{-1}+2P)a_{-2}
\nonumber\\
&\qquad  -\left(\mathtt{b}+\mathtt{b}^{-1}+2P\right)
 \left(\mathtt{b}+2\mathtt{b}^{-1}+2P\right)a_{-1}^2  
\nonumber\\
&\qquad + 2i(\mathtt{b}+2\mathtt{b}^{-1}+2P)a_{-1}L_{-1} - \mathtt{b}^{-1}(\mathtt{b}+\mathtt{b}^{-1}+2P)L_{-2}  + L_{-1}^2\Big)|P\rangle~,
\\[1mm]
|P;{\tiny\yng(2)},\emptyset\rangle &= \Big( -i\mathtt{b}
 (\mathtt{b}+\mathtt{b}^{-1}+2P)(2\mathtt{b}+\mathtt{b}^{-1}+2P)a_{-2} 
\nonumber\\
&\qquad - (\mathtt{b}+\mathtt{b}^{-1}+2P)(2\mathtt{b}+\mathtt{b}^{-1}+2P)a_{-1}^2 
\nonumber\\
&\qquad + 2i(2\mathtt{b}+\mathtt{b}^{-1}+2P)a_{-1}L_{-1}- \mathtt{b}(\mathtt{b}+\mathtt{b}^{-1}+2P)L_{-2} + L_{-1}^2 \Big)|P\rangle~,
\\[1mm]
|P;{\tiny\yng(1)},{\tiny\yng(1)}\rangle &=\Big(-i(\mathtt{b}+\mathtt{b}^{-1})a_{-2} - (\mathtt{b}^2
 + \mathtt{b}^{-2} +1-4P^2)a_{-1}^2  
\nonumber\\
&\qquad + 2i(\mathtt{b}+\mathtt{b}^{-1})a_{-1}L_{-1} - L_{-2}
  + L_{-1}^2 \Big)|P\rangle~,
\\[1mm]
|P;{\tiny\yng(1,1,1)},\emptyset\rangle &=
 \Big(-2i\mathtt{b}^{-2}(\mathtt{b}+\mathtt{b}^{-1}+2P)(\mathtt{b}+2\mathtt{b}^{-1}+2P)(\mathtt{b}+3\mathtt{b}^{-1}+2P)a_{-3}
\nonumber\\
&\qquad  -3\mathtt{b}^{-1}(\mathtt{b}+\mathtt{b}^{-1}+2P)(\mathtt{b}+2\mathtt{b}^{-1}+2P)(\mathtt{b}+3\mathtt{b}^{-1}+2P)a_{-2}a_{-1} 
\nonumber\\
&\qquad + i(\mathtt{b}+\mathtt{b}^{-1}+2P)(\mathtt{b}+2\mathtt{b}^{-1}+2P)(\mathtt{b}+3\mathtt{b}^{-1}+2P)a_{-1}^3
\nonumber\\
&\qquad + 3i\mathtt{b}^{-1}(\mathtt{b}+2\mathtt{b}^{-1}+2P)(\mathtt{b}+3\mathtt{b}^{-1}+2P)a_{-2}L_{-1} 
\nonumber\\
&\qquad + 3(\mathtt{b}+2\mathtt{b}^{-1}+2P)(\mathtt{b}+3\mathtt{b}^{-1}+2P)a_{-1}^2L_{-1} 
\nonumber\\
&\qquad +3i\mathtt{b}^{-1}(\mathtt{b}+\mathtt{b}^{-1}+2P)(\mathtt{b}+3\mathtt{b}^{-1}+2P)a_{-1}L_{-2}
\nonumber\\
&\qquad - 3i(\mathtt{b}+3\mathtt{b}^{-1}+2P)a_{-1}L_{-1}^2  - \mathtt{b}^{-2}(\mathtt{b}+4\mathtt{b}^{-1}+4P)(\mathtt{b}+\mathtt{b}^{-1}+2P)L_{-3}
\nonumber\\
&\qquad +
 \mathtt{b}^{-1}(3\mathtt{b}+5\mathtt{b}^{-1}+6P)L_{-2}L_{-1}-L_{-1}^3\Big)|P\rangle~,
\end{align}
\begin{align}
|P;{\tiny \yng(2,1)},\emptyset\rangle &= \Big(
 -i(\mathtt{b}+\mathtt{b}^{-1}+2P)(\mathtt{b}+2\mathtt{b}^{-1}+2P)(2\mathtt{b}+\mathtt{b}^{-1}+2P)a_{-3} 
\nonumber\\
& \qquad - (\mathtt{b}+\mathtt{b}^{-1}) (\mathtt{b}+\mathtt{b}^{-1} +
 2P)(\mathtt{b}+2\mathtt{b}^{-1}+2P)(2\mathtt{b}+\mathtt{b}^{-1}+2P)a_{-2}a_{-1}
\nonumber\\
&\qquad +i(\mathtt{b}+\mathtt{b}^{-1}+2P)(\mathtt{b}+2\mathtt{b}^{-1}+2P)(2\mathtt{b}+\mathtt{b}^{-1}+2P)a_{-1}^3
\nonumber\\
&\qquad + i(\mathtt{b}+\mathtt{b}^{-1})(\mathtt{b}+2\mathtt{b}^{-1}+2P)(2\mathtt{b}+\mathtt{b}^{-1}+2P)a_{-2}L_{-1} 
\nonumber\\
&\qquad + 3(\mathtt{b}+2\mathtt{b}^{-1}+2P)(2\mathtt{b}+\mathtt{b}^{-1}+2P)a_{-1}^2L_{-1}
\nonumber\\
&\qquad + i(\mathtt{b}+\mathtt{b}^{-1}+2P)(\mathtt{b}^2+5+\mathtt{b}^{-2}+2(\mathtt{b}+\mathtt{b}^{-1})P)a_{-1}L_{-2}
\nonumber\\
&\qquad -i(5\mathtt{b}+5\mathtt{b}^{-1}+6P)a_{-1}L_{-1}^2 - (\mathtt{b}+\mathtt{b}^{-1}+2P)^2L_{-3}  
\nonumber\\
&\qquad +(\mathtt{b}^2+3+\mathtt{b}^{-2}+2(\mathtt{b}+\mathtt{b}^{-1})P)L_{-2}L_{-1} - L_{-1}^3 \Big)|P\rangle~,
\\[1mm]
|P;{\tiny \yng(3)},\emptyset\rangle &= \Big( -2i\mathtt{b}^2
 (\mathtt{b}+\mathtt{b}^{-1}+2P)(3\mathtt{b}+\mathtt{b}^{-1}+2P)(2\mathtt{b}+\mathtt{b}^{-1}+2P)a_{-3}
\nonumber\\
&\qquad - 3\mathtt{b}(\mathtt{b}+\mathtt{b}^{-1}+2P)(3\mathtt{b}+\mathtt{b}^{-1}+2P)(2\mathtt{b}+\mathtt{b}^{-1}+2P)a_{-2}a_{-1}
\nonumber\\
&\qquad +i(\mathtt{b}+\mathtt{b}^{-1}+2P)(3\mathtt{b}+\mathtt{b}^{-1}+2P)(2\mathtt{b}+\mathtt{b}^{-1}+2P)a_{-1}^3 
\nonumber\\
&\qquad + 3i\mathtt{b} (3\mathtt{b}+\mathtt{b}^{-1}+2P)(2\mathtt{b}+\mathtt{b}^{-1}+2P)a_{-2}L_{-1}
\nonumber\\
&\qquad + 3(3\mathtt{b}+\mathtt{b}^{-1}+2P)(2\mathtt{b}+\mathtt{b}^{-1}+2P)a_{-1}^2L_{-1}
\nonumber\\
&\qquad + 3i\mathtt{b}(\mathtt{b}+\mathtt{b}^{-1}+2P)(3\mathtt{b}+\mathtt{b}^{-1}+2P)a_{-1}L_{-2} 
\nonumber\\
&\qquad -3i(3\mathtt{b}+\mathtt{b}^{-1}+2P)a_{-1}L_{-1}^2 
 -\mathtt{b}^2(\mathtt{b}+\mathtt{b}^{-1}+2P)(4\mathtt{b}+\mathtt{b}^{-1}+4P)L_{-3}
\nonumber\\
&\qquad  +\mathtt{b}(5\mathtt{b}+3\mathtt{b}^{-1}+6P)L_{-2}L_{-1}-L_{-1}^3 \Big)|P\rangle~,
\\[1mm]
|P; {\tiny\yng(1,1)},{\tiny\yng(1)}\rangle &=
 \Big(-2i\mathtt{b}^{-1}(\mathtt{b}+\mathtt{b}^{-1})(\mathtt{b}+\mathtt{b}^{-1}+2P)a_{-3} 
\nonumber\\
&\qquad - \mathtt{b}^{-1}(\mathtt{b}+\mathtt{b}^{-1}+2P)(3\mathtt{b}^2+3+2\mathtt{b}^{-2}-2\mathtt{b}^{-1}P-4P^2)a_{-2}a_{-1}
\nonumber\\
&\qquad + i (\mathtt{b}+\mathtt{b}^{-1}+2P)(\mathtt{b}^2+1+2\mathtt{b}^{-2}-2\mathtt{b}^{-1}P-4P^2)a_{-1}^3 
\nonumber\\
&\qquad + i\mathtt{b}^{-1}(3\mathtt{b}^2+7+2\mathtt{b}^{-2}+(4\mathtt{b}+6\mathtt{b}^{-1})P+4P^2)a_{-2}L_{-1}
\nonumber\\
&\qquad +(3\mathtt{b}^2+7+6\mathtt{b}^{-2}+(4\mathtt{b}+2\mathtt{b}^{-1})P-4P^2)a_{-1}^2L_{-1}
\nonumber\\
&\qquad +i\mathtt{b}^{-1}(3\mathtt{b}+\mathtt{b}^{-1}-2P)(\mathtt{b}+\mathtt{b}^{-1}+2P)a_{-1}L_{-2} 
\nonumber\\
&\qquad- i (3\mathtt{b}+5\mathtt{b}^{-1}+2P)a_{-1}L_{-1}^2 -
 \mathtt{b}^{-1}(\mathtt{b}+\mathtt{b}^{-1}+2P)L_{-3}
\nonumber\\
&\qquad  + \mathtt{b}^{-1}(3\mathtt{b}+\mathtt{b}^{-1}+2P)L_{-2}L_{-1}- L_{-1}^3\Big)|P\rangle~,
\end{align}
\begin{align}
|P; {\tiny\yng(2)},{\tiny\yng(1)}\rangle &=
 \Big(-2i\mathtt{b}(\mathtt{b}+\mathtt{b}^{-1})(\mathtt{b}+\mathtt{b}^{-1}+2P)a_{-3} 
\nonumber\\
&\qquad - \mathtt{b}(\mathtt{b}+\mathtt{b}^{-1}+2P)(2\mathtt{b}^2+3+3\mathtt{b}^{-2}-2\mathtt{b}P -4P^2)a_{-2}a_{-1}
\nonumber\\
&\qquad +i(\mathtt{b}+\mathtt{b}^{-1}+2P)(2\mathtt{b}^2+1+\mathtt{b}^{-2}-2\mathtt{b}P-4P^2)a_{-1}^3
\nonumber\\
&\qquad + i\mathtt{b}(2\mathtt{b}^2+7+3\mathtt{b}^{-2}+(6\mathtt{b}+4\mathtt{b}^{-1})P + 4P^2)a_{-2}L_{-1}
\nonumber\\
&\qquad +(6\mathtt{b}^2 + 7 + 3\mathtt{b}^{-2} + (2\mathtt{b}+4\mathtt{b}^{-1})P - 4P^2)a_{-1}^2L_{-1}
\nonumber\\
&\qquad +i\mathtt{b}(\mathtt{b}+3\mathtt{b}^{-1}-2P)(\mathtt{b}+\mathtt{b}^{-1}+2P)a_{-1}L_{-2}
\nonumber\\
&\qquad - i(5\mathtt{b}+3\mathtt{b}^{-1}+2P)a_{-1}L_{-1}^2 - \mathtt{b}(\mathtt{b}+\mathtt{b}^{-1}+2P)L_{-3}
\nonumber\\
&\qquad  
+
 \mathtt{b}(\mathtt{b}+3\mathtt{b}^{-1}+2P)L_{-2}L_{-1}- L_{-1}^3\Big)|P\rangle~.
\end{align}
Note that there are also states obtained by exchanging two Young
diagrams in each of the above, whose expressions are simply obtained via the relation
$|P;Y_2,Y_1\rangle = |-P;Y_1,Y_2\rangle$.

\section{Prepotential with massive deformations}
\label{app:F2}

As discussed in Sec.~\ref{subsec:massive-def}, we have checked that the instanton part of the
prepotential of the $(A_3,A_3)$ theory is invariant under
\eqref{eq:T-massive}, up to terms affected by the $U(1)$-factor. Indeed, our computation shows that
\begin{align}
 \mathcal{F}_\text{inst}^{(A_3,A_3)} = \sum_{k=-1}^\infty\mathcal{F}_{2k}a^{-2k}~,
\end{align}
where
\begin{align}
 \mathcal{F}_{-2} &= q^2 + \frac{13}{8}q^4 + \frac{23}{6}q^6 +
 \frac{2701}{256}q^8 + \mathcal{O}(q^{10})~,
\\[1mm]
\mathcal{F}_{2} &= \frac{q}{2} M u_1 u_2+ \frac{q^2}{16} \big(-4 d_1^2 d_2^2
 M^2-8 d_1^2 M^2 m_2-d_1^2 u_2^2
\nonumber\\
&\qquad\qquad \qquad \qquad   -8 d_2^2
 M^2 m_1-d_2^2 u_1^2-16 M^2 m_1 m_2-2
 m_1 u_2^2-2 m_2 u_1^2\big)
\nonumber\\
&\quad +\frac{q^3}{24}
 \big(12 d_1^3 d_2 M^2-6 d1^2 d_2 M
 u_1+12 d_1 d_2^3 M^2-6 d_1 d_2^2 M
 u_2+16 d_1 d_2 M^3
\nonumber\\
&\qquad \qquad\qquad + 24 d_1 d_2 M^2
 m_1+24 d_1 d_2 M^2 m_2+3 d1 d_2
 u_1^2+3 d_1 d_2 u_2^2
\nonumber\\[2mm]
&\qquad \qquad \qquad -12 d1 M m_2
 u_2-12 d_2 M m_1 u_1+12 M u_1
 u_2\big) + \mathcal{O}(q^4)~,
\end{align}
\begin{align}
\mathcal{F}_{4} &=  \frac{3q^2}{64} \left(4 d_1^2 M^2 u_2^2+4 d_2^2 M^2 u_1^2+8 M^2 m_1 u_2^2+8 M^2 m_2 u_1^2+u_1^2 u_2^2\right)
\nonumber\\
&\qquad + \frac{q^3}{48} \Big(-12 d_1^2 d_2^2 M u_1 u_2+20 d_1^2 d_2 M^3
 u_1-24 d_1^2 M m_2 u_1 u_2 + 20 d_1 d_2^2 M^3 u_2
\nonumber\\
&\qquad \qquad \qquad  -18 d_1 d_2 M^2 u_1^2-18 d_1d_2 M^2 u_2^2+40 d_1
 M^3 m_2 u_2+5 d_1 M u_2^3 -24 d_2^2 M m_1 u_1 u_2
\nonumber\\
&\qquad \qquad \qquad  + 40 d_2 M^3 m_1 u_1+5 d_2 M u_1^3-48 M m_1 m_2
 u_1 u_2\Big) + \mathcal{O}(q^4)~.
\end{align}
 One can explicitly check
that these expressions are all invariant under
\eqref{eq:T-massive}.\footnote{Note that $\mathcal{F}_0$ depends on the
$U(1)$-factor that we have not identified.}

%\subsection{Manifestly invariant expression}

As mentioned at the end of Sec.~\ref{subsec:massive-def}, imposing the
invariance under \eqref{eq:T-massive} constrains the possible
form of $\mathcal{F}_{k>0}$ in \eqref{eq:massiveF}.
In particular, when combined with the trivial symmetry of the theory
under exchanging two $(A_1,D_4)$ sectors, it determines $\mathcal{F}_k$
up to undetermined $T$-invariant functions. In the rest of this appendix, we
explicitly show this for $\mathcal{F}_2$.

To that end, it is useful for us below to re-express \eqref{eq:T-massive} in terms of the IR gauge coupling
$\tau_\text{IR}$ instead of the UV gauge coupling $q$, i.e.,
\begin{align}
 \tau_\text{IR}\to \tau_\text{IR}+ 1~,\qquad d_1 \to
 \frac{\theta_3^2}{\theta_4^2}d_1- \frac{\theta_2^2}{\theta_4^2}d_2~, \qquad d_2 \to
 id_2~,\qquad m_2\to -m_2~,\qquad u_2 \to iu_2~,
\label{eq:constraint1}
\end{align}
where we used the shorthands $\theta_2 \equiv \theta_2(q_\text{IR}^2),
\theta_3 \equiv \theta_3(q_\text{IR}^2)$ and $\theta_4\equiv \theta_4(q_\text{IR}^2)$.
Note that $\tau_\text{IR}\to \tau_\text{IR} +1$ implies $\theta_3(q_\text{IR}^2) \leftrightarrow
 \theta_4(q_\text{IR}^2)$ and $\theta_2(q_\text{IR}^2) \to
 e^{\frac{\pi i}{4}}\theta_2(q_\text{IR}^2)$.
%We also denote by $\tilde{q}_\text{IR}$ the image of $q_\text{IR}$
%under the $T$ transformation.

The $(A_3,A_3)$ theory also has a trivial
symmetry under the permutation of two $(A_1,D_4)$ theories in
the quiver description in Fig.~\ref{fig:quiver1}. This permutation symmetry
implies that $\mathcal{F}_{k>0}$ must also be invariant under 
\begin{align}
 d_1 \longleftrightarrow d_2~,\qquad m_1\longleftrightarrow m_2~,\qquad
 u_1\longleftrightarrow u_2~.
\label{eq:constraint2}
\end{align}
Below, we show that demanding the invariance of $\mathcal{F}_{k>0}$
under \eqref{eq:constraint1} and \eqref{eq:constraint2} strongly constrains the possible form
of $\mathcal{F}_{k>0}$, especially focusing on $\mathcal{F}_2$.

For that, first note that $\mathcal{F}_2$ is a meromorphic function of
$d_i,m_i,u_i,M$ and $\tau_\text{IR}$ such that $\mathcal{F}_2=0$ when
$d_i=m_i=u_i=M=0$. Moreover, since the prepotential is of dimension two,
$\mathcal{F}_2$ is of dimension four. These mean that $\mathcal{F}_2$
is a polynomial of $d_i,m_i,u_i$ and $M$ whose coefficients depend on $\tau_\text{IR}$.
The invariance under \eqref{eq:constraint1} and
\eqref{eq:constraint2} implies that these coefficients are
correlated in a highly non-trivial way. For example, let us consider the
possible term 
\begin{align}
 g(q_\text{IR})m_1u_1u_2
\label{eq:possibility1}
\end{align}
where $g(q_\text{IR})$ is an unknown function of $q_\text{IR}$.
 Since it is of dimension
four, this term might appear in $\mathcal{F}_2$. The constraint
 \eqref{eq:constraint2} then implies that \eqref{eq:possibility1} must be
 accompanied with $g(q_\text{IR})m_2u_1u_2$. Therefore,
 \eqref{eq:possibility1} can appear in $\mathcal{F}_2$ only as a part of
\begin{align}
 g(q_\text{IR})\left(m_1 + m_2\right)u_1u_2~.
\end{align}
However, the $T$-invariance \eqref{eq:constraint1} now requires
\begin{align}
ig(\tilde{q}_\text{IR})(m_1-m_2)u_1u_2 = g(q_\text{IR})(m_1+m_2)u_1u_2~,
\end{align}
where $\tilde{q}_\text{IR}$ is the image of $q_\text{IR}$ under the
$T$-transformation. For this equality to hold as a function of
$q_\text{IR},\,m_i$ and $u_i$, we need $g(q_\text{IR}) = 0$. Hence, the invariance under
the $T$-transformation together with the permutation symmetry \eqref{eq:constraint2} prohibits the term
proportional to $m_1u_1u_2$ or $m_2u_1u_2$ in $\mathcal{F}_2$.

% The invariance under \eqref{eq:constraint1} and \eqref{eq:constraint2}
% also constrains the coefficients of non-vanishing terms in $\mathcal{F}_2$. To see this,
% let us next consider the terms
% \begin{align}
% h_1(q_\text{IR}) (d_1^2 + d_2^2)u_1u_2 + h_2(q_\text{IR})d_1d_2u_1u_2~,
% \label{eq:ex2}
% \end{align}
% where $h_1(q_\text{IR})$ and $h_2(q_\text{IR})$ are unknown functions of
% $q_\text{IR}$. The coefficients are already chosen so that
% \eqref{eq:ex2} is invariant under \eqref{eq:constraint2}. The invariance
% under \eqref{eq:constraint1} then requires 
% $h_1(q_\text{IR}) = \frac{\theta_2^2}{\theta_4^4}H(q_\text{IR})$ and
% $h_2(q_\text{IR}) = 2\frac{\theta_3^2}{\theta_4^4}H(q_\text{IR})$, where
% $H(q_\text{IR})$ is a $T$-invariant function.
% %$h_2(q_\text{IR}) =
% %2h_1(q_\text{IR})\frac{\theta_3^2}{\theta_2^2}$. 
% Therefore, 
% terms of the form \eqref{eq:ex2} can
% appear in $\mathcal{F}_{2}$ only as 
% \begin{align}
%  H(q_\text{IR})\left(\left(d_1^2 + d_2^2\right)\frac{\theta_2^2}{\theta_4^4} + 2d_1d_2\frac{\theta_3^2}{\theta_4^4}\right)u_1u_2~.
% \end{align}

% As seen above, the invariance under the $T$ transformation
% \eqref{eq:constraint1} and the permutation \eqref{eq:constraint2} strongly constrains the possible form of
% $\mathcal{F}_{k>0}$. In appendix \ref{app:F2}, we solve these
% constraints to find an explicit expression for $\mathcal{F}_2$ which is
% manifestly invariant under \eqref{eq:constraint1} and
% \eqref{eq:constraint2}.

The invariance under \eqref{eq:constraint1} and \eqref{eq:constraint2}
also constrains the coefficients of non-vanishing terms in
$\mathcal{F}_2$. Indeed, we find that the most general expression for
$\mathcal{F}_2$ that is invariant under \eqref{eq:constraint1} and
\eqref{eq:constraint2} is written in terms of only 54 coefficients
$h_1(q_\text{Ir}),\cdots, h_{54}(q_\text{IR})$ as follows.
%
% Note that the above discussions also apply to  $\mathcal{F}_{k}$
% for all $k>0$. Therefore, the $T$-invariance and the permutation
% symmetry determine all $\mathcal{F}_{k>0}$ in terms of $T$-invariant
% functions. These $T$-invariant functions are then expected to be fixed so
% that the prepotential behaves as in \eqref{eq:Legendre} under the $S$
% transformation. We leave the detailed study of these $T$-invariant functions for future work.
% ***************************************
%
% In this appendix, we show an expression for $\mathcal{F}_2$ in
% \eqref{eq:inst-F} which is manifestly invariant under the $T$
% transformation \eqref{eq:T-massive} and the permutation symmetry. Note
% first that, since $\mathcal{F}_2$ is of dimension four, it has only
% finite number of terms. Some of the terms are already discussed in Sec.~\ref{subsec:T-perm}. Here we use $h_k(q_{\text{IR}})$ as coefficients for these terms, which is the unknown $T$-invariant function $h_k(q_{\text{IR}})$. Thus, as with Sec.~\ref{subsec:T-perm}, the explicit expression for $\mathcal{F}_2$ is written as
\begin{align}
\begin{aligned}
	&\mathcal{F}_2(q,M,\{m_i\},\{u_i\},\{d_i\})\\
	&=h_1(q_{\text{IR}})M^4+h_2(q_{\text{IR}})M^2(m_1^2+m_2^2)+h_3(q_{\text{IR}})M^2m_1m_2\frac{\theta_2^4}{\theta_3^2\theta_4^2}+h_4(q_{\text{IR}})(m_1^4+m_2^4)\\
	&\qquad+h_5(q_{\text{IR}}) m_1^2m_2^2+h_6(q_{\text{IR}})m_1m_2(m_1^2+m_2^2)\frac{\theta_2^4}{\theta_3^2\theta_4^2}\\
	&\quad+h_7(q_{\text{IR}})Mu_1u_2\frac{\theta_2^2}{\theta_3\theta_4}+h_8(q_{\text{IR}})(m_1u_1^2+m_2u_2^2)+h_9(q_{\text{IR}}) (m_2u_1^2+m_1u_2^2)\frac{\theta_2^4}{\theta_3^2\theta_4^2}\\
	&\quad+h_{10}(q_{\text{IR}})M^2B_2(u_i,d_i)+h_{11}(q_{\text{IR}})m_1m_2B_2(u_i,d_i)\frac{\theta_2^4}{\theta_3^2\theta_4^2}+h_{12}(q_{\text{IR}})MB_3^{(1)}(m_i,u_i,d_i)\\
	&\qquad+h_{13}(q_{\text{IR}})MB_3^{(2)}(m_i,u_i,d_i)+h_{14}(q_{\text{IR}})B_4^{(1)}(m_i,u_i,d_i)+h_{15}(q_{\text{IR}})B_4^{(2)}(m_i,u_i,d_i)\\
	&\quad+h_{16}(q_{\text{IR}})M^3D_1(d_i)+h_{17}(q_{\text{IR}})M(m_1^2+m_2^2)D_1(d_i)+h_{18}(q_{\text{IR}})Mm_1m_2D_1(d_i)\frac{\theta_2^4}{\theta_3^2\theta_4^2}\\
	&\qquad+h_{19}(q_{\text{IR}})M^2E_2^{(1)}(m_i,d_i)+h_{20}(q_{\text{IR}})M^2E_2^{(2)}(m_i,d_i)+h_{21}(q_{\text{IR}})m_1m_2E_2^{(1)}(m_i,d_i)\frac{\theta_2^4}{\theta_3^2\theta_4^2}\\
	&\qquad+h_{22}(q_{\text{IR}})m_1m_2E_2^{(2)}(m_i,d_i)\frac{\theta_2^4}{\theta_3^2\theta_4^2}+h_{23}(q_{\text{IR}})E_4^{(1)}(m_i,d_i)+h_{24}(q_{\text{IR}})E_4^{(2)}(m_i,d_i)\\
	&\quad+h_{25}(q_{\text{IR}})u_1u_2D_1(d_i)\frac{\theta_2^2}{\theta_3\theta_4}+h_{26}(q_{\text{IR}})F_4^{(1)}(u_i,d_i)+h_{27}(q_{\text{IR}})F_4^{(2)}(u_i,d_i)\\
	&\quad+h_{28}(q_{\text{IR}})MG_3^{(1)}(u_i,d_i)+h_{29}(q_{\text{IR}})MG_3^{(2)}(u_i,d_i)+h_{30}(q_{\text{IR}})G_4^{(1)}(m_i,u_i,d_i)\\
	&\qquad+h_{31}(q_{\text{IR}})G_4^{(2)}(m_i,u_i,d_i)+h_{32}(q_{\text{IR}})G_4^{(3)}(m_i,u_i,d_i)+h_{33}(q_{\text{IR}})G_4^{(4)}(m_i,u_i,d_i)\\
	&\quad+h_{34}(q_{\text{IR}})M^2(D_1(d_i))^2+h_{35}(q_{\text{IR}})m_1m_2(D_1(d_i))^2\frac{\theta_2^4}{\theta_3^2\theta_4^2}+h_{36}(q_{\text{IR}})M^2D_2(d_i)\\
	&\qquad+h_{37}(q_{\text{IR}})m_1m_2D_2(d_i)\frac{\theta_2^4}{\theta_3^2\theta_4^2}+h_{38}(q_{\text{IR}}) (m_1^2+m_2^2)D_2(d_i)+h_{39}(q_{\text{IR}})MH_3^{(1)}(m_i,d_i)\\
	&\qquad+h_{40}(q_{\text{IR}})MH_3^{(2)}(m_i,d_i)+h_{41}(q_{\text{IR}})H_4^{(1)}(m_i,d_i)+h_{42}(q_{\text{IR}})H_4^{(2)}(m_i,d_i)\\
	&\quad+h_{43}(q_{\text{IR}})I_4^{(1)}(u_i,d_i)+h_{44}(q_{\text{IR}})I_4^{(2)}(u_i,d_i)+h_{45}(q_{\text{IR}})I_4^{(3)}(u_i,d_i)\\
	&\quad+h_{46}(q_{\text{IR}})M(D_1(d_i))^3+h_{47}(q_{\text{IR}})MD_1(d_i)D_2(d_i)+h_{48}(q_{\text{IR}})J_4^{(1)}(m_i,d_i)\\
	&\qquad+h_{49}(q_{\text{IR}})J_4^{(2)}(m_i,d_i)+h_{50}(q_{\text{IR}})J_4^{(3)}(m_i,d_i)+h_{51}(q_{\text{IR}})J_4^{(4)}(m_i,d_i)\\
	&\quad+h_{52}(q_{\text{IR}})(D_1(d_i))^4+h_{53}(q_{\text{IR}})(D_1(d_i))^2D_2(d_i)+h_{54}(q_{\text{IR}}) (D_2(d_i))^2
\end{aligned}
\end{align}
with the following building blocks
\footnotesize
\begin{align}
	&B_2(u_i,d_i)=\bigg((u_1d_1+u_2d_2)+(u_2d_1+u_1d_2)\frac{\theta_3^2-\theta_4^2}{\theta_2^2}\bigg)\frac{\theta_3}{\theta_4}~,\\
	&B_3^{(1)}(m_i,u_i,d_i)=\bigg((m_1u_1d_1+m_2u_2d_2)+(m_2u_2d_1+m_1u_1d_2)\frac{\theta_3^2+\theta_4^2}{\theta_2^2}\bigg)\frac{\theta_3}{\theta_4}~,\\
	&B_3^{(2)}(m_i,u_i,d_i)=\bigg((m_2u_1d_1+m_1u_2d_2)+(m_1u_2d_1+m_2u_1d_2)\frac{\theta_3^2+\theta_4^2}{\theta_2^2}\bigg)\frac{\theta_2^4}{\theta_3\theta_4^3}~,\\
	&B_4^{(1)}(m_i,u_i,d_i)=\bigg((m_1^2u_1d_1+m_2^2u_2d_2)+(m_2^2u_2d_1+m_1^2u_1d_2)\frac{\theta_3^2-\theta_4^2}{\theta_2^2}\bigg)\frac{\theta_3}{\theta_4}~,\\
	&B_4^{(2)}(m_i,u_i,d_i)=\bigg((m_2^2u_1d_1+m_1^2u_2d_2)+(m_1^2u_2d_1+m_2^2u_1d_2)\frac{\theta_3^2-\theta_4^2}{\theta_2^2}\bigg)\frac{\theta_3}{\theta_4}~,\\
	&D_1(d_i)=\bigg((d_1^2+d_2^2)+2d_1d_2\frac{\theta_3^2}{\theta_2^2}\bigg)\frac{\theta_3^2}{\theta_4^2}~,\\
	&D_2(d_i)=\bigg((d_1^3d_2+d_1d_2^3)+d_1^2d_2^2\frac{\theta_3^4+\theta_2^4}{\theta_3^2\theta_2^2}\bigg)\frac{\theta_3^4}{\theta_4^2\theta_2^2}~,\\
	&E_2^{(1)}(m_i,d_i)=\bigg( (m_1d_1^2+m_2d_2^2)+(m_1+m_2)d_1d_2\frac{\theta_2^2}{\theta_3^2}\bigg)\frac{\theta_3^2}{\theta_4^2}~,\\
	&E_2^{(2)}(m_i,d_i)=\bigg( (m_2d_1^2+m_1d_2^2)+(m_1+m_2)d_1d_2\frac{\theta_2^2}{\theta_3^2}\bigg)\frac{\theta_2^4}{\theta_4^4}~,\\
	&E_4^{(1)}(m_i,d_i)=\bigg( (m_1^3d_1^2+m_2^3d_2^2)+(m_1^3+m_2^3)d_1d_2\frac{\theta_2^2}{\theta_3^2}\bigg)\frac{\theta_3^2}{\theta_4^2}~,\\
	&E_4^{(2)}(m_i,d_i)=\bigg( (m_2^3d_1^2+m_1^3d_2^2)+(m_1^3+m_2^3)d_1d_2\frac{\theta_2^2}{\theta_3^2}\bigg)\frac{\theta_2^4}{\theta_4^4}~,\\
	&F_4^{(1)}(u_i,d_i)=\bigg( (u_1^2d_1^2+ u_2^2d_2^2)+d_1d_2(u_1^2+u_2^2)\frac{\theta_2^2}{\theta_3^2} \bigg)\frac{\theta_3^2}{\theta_4^2}~,\\
	&F_4^{(2)}(u_i,d_i)=\bigg( (u_2^2d_1^2+ u_1^2d_2^2)+d_1d_2(u_1^2+u_2^2)\frac{\theta_2^2}{\theta_3^2} \bigg)\frac{\theta_2^4}{\theta_4^4}~,\\
	&G_3^{(1)}(u_i,d_i)=\bigg( (u_1d_1^3+u_2d_2^3)+d_1d_2\Big((u_1d_1+u_2d_2)\frac{\theta_3^4+\theta_2^4+\theta_3^2\theta_4^2}{\theta_3^2\theta_2^2}+(u_2d_1+u_1d_2)\frac{\theta_3^2+\theta_4^2}{\theta_3^2}\Big) \bigg)\frac{\theta_3^3}{\theta_4^3}~,\\
	&G_3^{(2)}(u_i,d_i)=\bigg( (u_2d_1^3+u_1d_2^3)+d_1d_2\Big((u_1d_1+u_2d_2)\frac{\theta_3^2-\theta_4^2}{\theta_3^2}+(u_2d_1+u_1d_2)\frac{\theta_3^4+\theta_2^4-\theta_3^2\theta_4^2}{\theta_3^2\theta_2^2}\Big) \bigg)\frac{\theta_3^2\theta_2^2}{\theta_4^4}~,\\
	&G_4^{(1)}(m_i,u_i,d_i)=\bigg( (m_1u_1d_1^3+m_2u_2d_2^3)\nonumber\\
	&\qquad\qquad\qquad+d_1d_2\Big((m_1u_1d_1+m_2u_2d_2)\frac{\theta_3^4+\theta_2^4-\theta_3^2\theta_4^2}{\theta_3^2\theta_2^2}+(m_2u_2d_1+m_1u_1d_2)\frac{\theta_3^2-\theta_4^2}{\theta_3^2}\Big) \bigg)\frac{\theta_3^3}{\theta_4^3}~,\\
	&G_4^{(2)}(m_i,u_i,d_i)=\bigg( (m_2u_1d_1^3+m_1u_2d_2^3)\nonumber\\
	&\qquad\qquad\qquad+d_1d_2\Big((m_2u_1d_1+m_1u_2d_2)\frac{\theta_3^4+\theta_2^4-\theta_3^2\theta_4^2}{\theta_3^2\theta_2^2}+(m_1u_2d_1+m_2u_1d_2)\frac{\theta_3^2-\theta_4^2}{\theta_3^2}\Big)
 \bigg)\frac{\theta_3\theta_2^4}{\theta_4^5}~,
\end{align}
\begin{align}
	&G_4^{(3)}(m_i,u_i,d_i)=\bigg( (m_1u_2d_1^3+m_2u_1d_2^3)\nonumber\\
	&\qquad\qquad\qquad+d_1d_2\Big((m_2u_1d_1+m_1u_2d_2)\frac{\theta_3^2+\theta_4^2}{\theta_3^2}+(m_1u_2d_1+m_2u_1d_2)\frac{\theta_3^4+\theta_2^4+\theta_3^2\theta_4^2}{\theta_3^2\theta_2^2} \Big) \bigg)\frac{\theta_3^2\theta_2^2}{\theta_4^4}~,\\
	&G_4^{(4)}(m_i,u_i,d_i)=\bigg( (m_2u_2d_1^3+m_1u_1d_2^3)\nonumber\\
	&\qquad\qquad\qquad+d_1d_2\Big((m_1u_1d_1+m_2u_2d_2)\frac{\theta_3^2+\theta_4^2}{\theta_3^2}+(m_2u_2d_1+m_1u_1d_2)\frac{\theta_3^4+\theta_2^4+\theta_3^2\theta_4^2}{\theta_3^2\theta_2^2}
 \Big) \bigg)\frac{\theta_3^4}{\theta_4^2\theta_2^2}~,
\\
 	&H_3^{(1)}(m_i,d_i)=\bigg( (m_1d_1^4+m_2d_2^4)\nonumber\\
	&\qquad\qquad\qquad+d_1d_2\Big((m_1d_1^2+m_2d_2^2)\frac{2\theta_3^4+\theta_2^4}{\theta_2^2\theta_3^2}+(m_2d_1^2+m_1d_2^2)\frac{\theta_2^2}{\theta_3^2}\Big)+3(m_1+m_2)d_1^2d_2^2 \bigg)\frac{\theta_3^4}{\theta_4^4}~,\\
	&H_3^{(2)}(m_i,d_i)=\bigg( (m_2d_1^4+m_1d_2^4)\nonumber\\
	&\qquad\qquad\qquad+d_1d_2\Big((m_1d_1^2+m_2d_2^2)\frac{\theta_2^2}{\theta_3^2}+(m_2d_1^2+m_1d_2^2)\frac{2\theta_3^4+\theta_2^4}{\theta_2^2\theta_3^2}\Big)+3(m_1+m_2)d_1^2d_2^2 \bigg)\frac{\theta_3^2\theta_2^4}{\theta_4^6}~,\\
	&H_4^{(1)}(m_i,d_i)=\bigg( (m_1^2d_1^4+m_2^2d_2^4)-2d_1d_2(m_2^2d_1^2+m_1^2d_2^2)\frac{\theta_2^2}{\theta_3^2}-(m_1^2+m_2^2)d_1^2d_2^2\frac{2\theta_3^4+\theta_2^4}{\theta_3^4} \bigg)\frac{\theta_3^4}{\theta_4^4}~,\\
	&H_4^{(2)}(m_i,d_i)=\bigg( (m_2^2d_1^4+m_1^2d_2^4)+2d_1d_2(m_2^2d_1^2+m_1^2d_2^2)\frac{\theta_2^2}{\theta_3^2}+(m_1^2+m_2^2)d_1^2d_2^2\frac{\theta_2^4}{\theta_3^4} \bigg)\frac{\theta_3^4}{\theta_4^4}~,\\
	&I_4^{(1)}(u_i,d_i)=\bigg( u_1d_1^5+u_2d_2^5+d_1d_2(u_2d_1^3+u_1d_2^3)\frac{-3\theta_2^8+2\theta_4^6(\theta_3^2-\theta_4^2)+\theta_4^2\theta_2^4(4\theta_3^2-5\theta_4^2)}{\theta_3^4\theta_2^4} \nonumber\\
	&\qquad\qquad\qquad+d_1^2d_2^2\Big((u_2d_1+u_1d_2)\frac{-8\theta_2^8(\theta_3^2-\theta_4^2)-2\theta_4^8(\theta_3^2-\theta_4^2)-\theta_4^4\theta_2^4(8\theta_3^2-9\theta_4^2)}{\theta_3^4\theta_2^6}\nonumber\\
	&\qquad\qquad\qquad+(u_1d_1+u_2d_2)\frac{-6\theta_2^8+4\theta_4^2\theta_2^4(\theta_3^2-2\theta_4^2)+3\theta_4^6(\theta_3^2-\theta_4^2)}{\theta_3^4\theta_2^4}\Big) \bigg)\frac{\theta_3^5}{\theta_4^5}~,\\
	&I_4^{(2)}(u_i,d_i)=\bigg( u_2d_1^5+u_1d_2^5+d_1d_2(u_2d_1^3+u_1d_2^3)\frac{\theta_4^4(\theta_3^2+\theta_4^2)+\theta_2^4(3\theta_3^2+\theta_4^2)}{\theta_3^4\theta_2^2}\nonumber\\
	&\qquad\qquad\qquad+d_1^2d_2^2\Big((u_1d_1+u_2d_2)\frac{\theta_2^2(\theta_3^2+\theta_4^2)}{\theta_3^4}+(u_2d_1+u_1d_2)\frac{3\theta_2^4+2\theta_4^2(\theta_3^2+\theta_4^2)}{\theta_3^4} \Big) \bigg)\frac{\theta_3^4\theta_2^4}{\theta_4^6}~,\\
	&I_4^{(3)}(u_i,d_i)=\bigg( d_1d_2(u_1d_1^3+u_2d_2^3)+d_1^2d_2^2\Big((u_1d_1+u_2d_2)\frac{\theta_4^4(2\theta_3^2-\theta_4^2)+\theta_2^4(3\theta_3^2-\theta_4^2)}{\theta_3^4\theta_2^2}\nonumber\\
	&\qquad\qquad\qquad+(u_2d_1+u_1d_2)\frac{3\theta_2^8-\theta_4^6(\theta_3^2-\theta_4^2)-\theta_4^2\theta_2^4(2\theta_3^2+4\theta_4^2)}{\theta_3^4\theta_2^4} \Big) \bigg)\frac{\theta_3^5}{\theta_4^3\theta_2^2}~,\\
	&J_4^{(1)}(m_i,d_i)=\bigg( (m_1d_1^6+m_2d_2^6) +3d_1d_2(m_1d_1^4+m_2d_2^4)\frac{\theta_2^2}{\theta_3^2} \nonumber\\
	&\qquad\qquad\qquad\qquad\qquad+3d_1^2d_2^2(m_1d_1^2+m_2d_2^2)\frac{\theta_2^4}{\theta_3^4}+(m_1+m_2)d_1^3d_2^3\frac{\theta_2^6}{\theta_3^6} \bigg)\frac{\theta_3^6}{\theta_4^6}~,\\
	&J_4^{(2)}(m_i,d_i)=\bigg( (m_2d_1^6+m_1d_2^6) +3d_1d_2(m_2d_1^4+m_1d_2^4)\frac{\theta_2^2}{\theta_3^2} \nonumber\\
	&\qquad\qquad\qquad\qquad\qquad+3d_1^2d_2^2(m_2d_1^2+m_1d_2^2)\frac{\theta_2^4}{\theta_3^4}+(m_1+m_2)d_1^3d_2^3\frac{\theta_2^6}{\theta_3^6}
 \bigg)\frac{\theta_3^4\theta_2^4}{\theta_4^8}~,
\end{align}
\begin{align}
	&J_4^{(3)}(m_i,d_i)=\bigg( d_1d_2(m_1d_1^4+m_2d_2^4)+d_1^2d_2^2\Big((m_1d_1^2+m_2d_2^2)\frac{\theta_3^4+2\theta_2^4}{\theta_3^2\theta_2^2}+(m_2d_1^2+m_1d_2^2)\frac{\theta_2^2}{\theta_3^2}\Big)\nonumber\\
	&\qquad\qquad\qquad\qquad\qquad+(m_1+m_2)d_1^3d_2^3\frac{\theta_3^8+\theta_2^8+\theta_4^4\theta_2^4}{\theta_3^8} \bigg)\frac{\theta_3^6}{\theta_4^4\theta_2^2}~,\\
	&J_4^{(4)}(m_i,d_i)=\bigg( d_1d_2(m_2d_1^4+m_1d_2^4)+d_1^2d_2^2\Big((m_1d_1^2+m_2d_2^2)\frac{\theta_2^2}{\theta_3^2}+(m_2d_1^2+m_1d_2^2)\frac{\theta_3^4+2\theta_2^4}{\theta_3^2\theta_2^2}\Big)\nonumber\\
	&\qquad\qquad\qquad\qquad\qquad+(m_1+m_2)d_1^3d_2^3\frac{\theta_3^8+\theta_2^8+\theta_4^4\theta_2^4}{\theta_3^8} \bigg)\frac{\theta_3^4\theta_2^2}{\theta_4^6}~,
\end{align}
\normalsize
where each 
%blocks are 
block is
 invariant under the $T$ transformation \eqref{eq:T-massive} and the
 permutation symmetry, and 
their 
subscripts 
%mean 
coincide with
%the 
their mass
dimensions.
% of them.

\bibliography{AGT}
\bibliographystyle{utphys}

\end{document}